\documentclass[twocolumn]{aastex63}
\usepackage{url}
\usepackage{threeparttable}
\usepackage{ulem}

\usepackage{natbib}
\usepackage{appendix}
\usepackage{amsmath}

\def\tcra{\textcolor{black}}
\def\tcrb{\textcolor{black}}
\def\tcrc{\textcolor{black}}
\def\tcrd{\textcolor{black}}
\def\tcre{\textcolor{black}}
\def\tcrf{\textcolor{black}}
\def\tcrg{\textcolor{black}}

\accepted{for publication in ApJ on 11 April, 2023}
\shorttitle{Kinematics of Extremely Metal-Poor Galaxies}
\shortauthors{Isobe et al.}
\graphicspath{{./}{figures/}}

\begin{document}

\title{EMPRESS. IX.\\
\tcra{
Extremely Metal-Poor Galaxies are Very Gas-Rich Dispersion-Dominated Systems:\\
Will JWST Witness Gaseous Turbulent High-$z$ Primordial Galaxies?
}
}

\author[0000-0001-7730-8634]{Yuki Isobe}
\affiliation{Institute for Cosmic Ray Research, The University of Tokyo, 5-1-5 Kashiwanoha, Kashiwa, Chiba 277-8582, Japan}
\affiliation{Department of Physics, Graduate School of Science, The University of Tokyo, 7-3-1 Hongo, Bunkyo, Tokyo 113-0033, Japan}

\author[0000-0002-1049-6658]{Masami Ouchi}
\affiliation{National Astronomical Observatory of Japan, 2-21-1 Osawa, Mitaka, Tokyo 181-8588, Japan}
\affiliation{Institute for Cosmic Ray Research, The University of Tokyo, 5-1-5 Kashiwanoha, Kashiwa, Chiba 277-8582, Japan}
\affiliation{Kavli Institute for the Physics and Mathematics of the Universe (WPI), University of Tokyo, Kashiwa, Chiba 277-8583, Japan}

\author[0000-0003-2965-5070]{Kimihiko Nakajima}
\affiliation{National Astronomical Observatory of Japan, 2-21-1 Osawa, Mitaka, Tokyo 181-8588, Japan}

\author[0000-0002-5443-0300]{Shinobu Ozaki}
\affiliation{National Astronomical Observatory of Japan, 2-21-1 Osawa, Mitaka, Tokyo 181-8588, Japan}

\author[0000-0003-0068-9920]{Nicolas F. Bouch{\'e}}
\affiliation{Univ Lyon, Univ Lyon1, ENS de Lyon, CNRS, Centre de Recherche Astrophysique de Lyon UMR5574, F-69230 Saint-Genis-Laval France}

\author[0000-0003-1173-8847]{John H. Wise}
\affiliation{Center for Relativistic Astrophysics, School of Physics, Georgia Institute of Technology, Atlanta, GA 30332, USA}

\author[0000-0002-5768-8235]{Yi Xu}
\affiliation{Institute for Cosmic Ray Research, The University of Tokyo, 5-1-5 Kashiwanoha, Kashiwa, Chiba 277-8582, Japan}
\affiliation{Department of Astronomy, Graduate School of Science, The University of Tokyo, 7-3-1 Hongo, Bunkyo, Tokyo 113-0033, Japan}

\author[0000-0002-6155-7166]{Eric Emsellem}
\affiliation{European Southern Observatory, Karl-Schwarzschild-Stra{\ss}e 2, 85748 Garching, Germany}
\affiliation{Univ Lyon, Univ Lyon1, ENS de Lyon, CNRS, Centre de Recherche Astrophysique de Lyon UMR5574, F-69230 Saint-Genis-Laval France}

\author[0000-0002-3801-434X]{Haruka Kusakabe} 
\affiliation{Observatoire de Gen{\'e}ve, Universit{\'e} de Gen{\'e}ve, 51 Ch. des Maillettes, 1290 Versoix, Switzerland}

\author[0000-0002-8996-7562]{Takashi Hattori}
\affiliation{Subaru Telescope, National Astronomical Observatory of Japan, National Institutes of Natural Sciences (NINS), 650 North A'ohoku Place, Hilo, HI 96720, USA}

\author[0000-0002-7402-5441]{Tohru Nagao}
\affiliation{Research Center for Space and Cosmic Evolution, Ehime
University, Bunkyo-cho 2-5, Matsuyama, Ehime 790-8577, Japan}

\author[0000-0001-6246-2866]{Gen Chiaki}
\affiliation{National Astronomical Observatory of Japan, 2-21-1 Osawa, Mitaka, Tokyo 181-8588, Japan}

\author[0000-0002-0547-3208]{Hajime Fukushima}
\affiliation{Center for Computational Sciences, University of Tsukuba, Ten-nodai, 1-1-1 Tsukuba, Ibaraki 305-8577, Japan}

\author[0000-0002-6047-430X]{Yuichi Harikane} 
\affiliation{Institute for Cosmic Ray Research, The University of Tokyo, 5-1-5 Kashiwanoha, Kashiwa, Chiba 277-8582, Japan}
\affiliation{Department of Physics and Astronomy, University College London, Gower Street, London WC1E 6BT, UK}

\author[0000-0002-8758-8139]{Kohei Hayashi}
\affiliation{National Institute of Technology, Ichinoseki College, Hagisho, Ichinoseki, 021-8511, Japan}
\affiliation{Astronomical Institute, Tohoku University, 6-3 Aoba, Aramaki, Aoba-ku, Sendai, Miyagi 980-8578, Japan}
\affiliation{Institute for Cosmic Ray Research, The University of Tokyo, 5-1-5 Kashiwanoha, Kashiwa, Chiba 277-8582, Japan}

\author[0000-0002-5661-033X]{Yutaka Hirai}
\affiliation{Department of Physics and Astronomy, University of Notre Dame, 225 Nieuwland Science Hall, Notre Dame, IN 46556, USA}
\affiliation{Astronomical Institute, Tohoku University, 6-3 Aoba, Aramaki, Aoba-ku, Sendai, Miyagi 980-8578, Japan}

\author[0000-0002-1418-3309]{Ji Hoon Kim}
\affiliation{Astronomy Program, Department of Physics and Astronomy, Seoul National University, 1 Gwanak-ro, Gwanak-gu, Seoul 08826, Republic of Korea}
\affiliation{SNU Astronomy Research Center, Seoul National University, 1 Gwanak-ro, Gwanak-gu, Seoul 08826, Republic of Korea}

\author[0000-0003-0695-4414]{Michael V. Maseda}
\affiliation{Department of Astronomy, University of Wisconsin-Madison, 475 N. Charter Street, Madison, WI 53706, USA}

\author[0000-0001-7457-8487]{Kentaro Nagamine}
\affiliation{Theoretical Astrophysics, Department of Earth \& Space Science, Graduate School of Science, Osaka University, 
1-1 Machikaneyama, Toyonaka, Osaka 560-0043, Japan}
\affiliation{Kavli Institute for the Physics and Mathematics of the Universe (WPI), University of Tokyo, Kashiwa, Chiba 277-8583, Japan}
\affiliation{Department of Physics \& Astronomy, University of Nevada, Las Vegas, 4505 S. Maryland Pkwy, Las Vegas, NV 89154-4002, USA}

\author{Takatoshi Shibuya}
\affiliation{Kitami Institute of Technology, 165 Koen-cho, Kitami, Hokkaido 090-8507, Japan}

\author[0000-0001-6958-7856]{Yuma Sugahara} 
\affiliation{National Astronomical Observatory of Japan, 2-21-1 Osawa, Mitaka, Tokyo 181-8588, Japan}
\affiliation{Waseda Research Institute for Science and Engineering, Faculty of Science and Engineering, Waseda University, 3-4-1, Okubo, Shinjuku, Tokyo 169-8555, Japan}

\author[0000-0002-1319-3433]{Hidenobu Yajima}
\affiliation{Center for Computational Sciences, University of Tsukuba, Ten-nodai, 1-1-1 Tsukuba, Ibaraki 305-8577, Japan}

\author[0000-0002-1005-4120]{Shohei Aoyama}
\affiliation{Institute of Management and Information Technologies, Chiba University, 1-33, Yayoi-cho, Inage-ward, Chiba, 263-8522, Japan}
\affiliation{Institute for Cosmic Ray Research, The University of Tokyo, 5-1-5 Kashiwanoha, Kashiwa, Chiba 277-8582, Japan}

\author[0000-0001-7201-5066]{Seiji Fujimoto} 
\affiliation{Cosmic DAWN Center}
\affiliation{Niels Bohr Institute, University of Copenhagen, Lyngbyvej2, DK-2100, Copenhagen, Denmark}
\affiliation{Research Institute for Science and Engineering, Waseda University, 3-4-1 Okubo, Shinjuku, Tokyo 169-8555, Japan}
\affiliation{National Astronomical Observatory of Japan, 2-21-1 Osawa, Mitaka, Tokyo 181-8588, Japan}
\affiliation{Institute for Cosmic Ray Research, The University of Tokyo, 5-1-5 Kashiwanoha, Kashiwa, Chiba 277-8582, Japan}

\author{Keita Fukushima}
\affiliation{Theoretical Astrophysics, Department of Earth \& Space Science, Graduate School of Science, Osaka University, 
1-1 Machikaneyama, Toyonaka, Osaka 560-0043, Japan}


\author{Shun Hatano}
\affiliation{Department of Astronomical Science, SOKENDAI (The Graduate University for Advanced Studies), Osawa 2-21-1, Mitaka, Tokyo, 181-8588, Japan}

\author[0000-0002-7779-8677]{Akio K. Inoue}
\affiliation{Waseda Research Institute for Science and Engineering, Faculty of Science and Engineering, Waseda University, 3-4-1, Okubo, Shinjuku, Tokyo 169-8555, Japan}
\affiliation{Department of Physics, School of Advanced Science and Engineering, Faculty of Science and Engineering, Waseda University, 3-4-1 Okubo, Shinjuku, Tokyo 169-8555, Japan}

\author{Tsuyoshi Ishigaki}
\affiliation{Department of Physical Science and Materials Engineering, Faculty of Science and Engineering, Iwate University \\
3-18-34 Ueda, Morioka, Iwate 020-8550, Japan}

\author{Masahiro Kawasaki}
\affiliation{Institute for Cosmic Ray Research, The University of Tokyo, 5-1-5 Kashiwanoha, Kashiwa, Chiba 277-8582, Japan}
\affiliation{Kavli Institute for the Physics and Mathematics of the Universe (WPI), University of Tokyo, Kashiwa, Chiba 277-8583, Japan}

\author[0000-0001-5780-1886]{Takashi Kojima}
\affiliation{Institute for Cosmic Ray Research, The University of Tokyo, 5-1-5 Kashiwanoha, Kashiwa, Chiba 277-8582, Japan}
\affiliation{Department of Physics, Graduate School of Science, The University of Tokyo, 7-3-1 Hongo, Bunkyo, Tokyo 113-0033, Japan}

\author[0000-0002-3852-6329]{Yutaka Komiyama} 
\affiliation{Department of Advanced Sciences, Faculty of Science and Engineering, Hosei University, 3-7-2 Kajino-cho, Koganei-shi, Tokyo 184-8584, Japan}

\author{Shuhei Koyama}
\affiliation{Institute of Astronomy, Graduate School of Science, The University of Tokyo, 2-21-1 Osawa, Mitaka, Tokyo 181-0015, Japan}

\author[0000-0002-0479-3699]{Yusei Koyama}
\affiliation{Subaru Telescope, National Astronomical Observatory of Japan, National Institutes of Natural Sciences (NINS), 650 North A'ohoku Place, Hilo, HI 96720, USA}
\affiliation{Department of Astronomical Science, SOKENDAI (The Graduate University for Advanced Studies), Osawa 2-21-1, Mitaka, Tokyo, 181-8588, Japan}

\author[0000-0003-1700-5740]{Chien-Hsiu Lee} 
\affiliation{W. M. Keck Observatory, Kamuela, HI 96743, USA}

\author{Akinori Matsumoto}
\affiliation{Institute for Cosmic Ray Research, The University of Tokyo, 5-1-5 Kashiwanoha, Kashiwa, Chiba 277-8582, Japan}
\affiliation{Department of Physics, Graduate School of Science, The University of Tokyo, 7-3-1 Hongo, Bunkyo, Tokyo 113-0033, Japan}

\author[0000-0003-4985-0201]{Ken Mawatari}
\affiliation{National Astronomical Observatory of Japan, 2-21-1 Osawa, Mitaka, Tokyo 181-8588, Japan}

\author[0000-0003-1169-1954]{Takashi J. Moriya}
\affiliation{National Astronomical Observatory of Japan, 2-21-1 Osawa, Mitaka, Tokyo 181-8588, Japan}
\affiliation{School of Physics and Astronomy, Faculty of Science, Monash University, Clayton, Victoria 3800, Australia}

\author{Kentaro Motohara}
\affiliation{National Astronomical Observatory of Japan, 2-21-1 Osawa, Mitaka, Tokyo 181-8588, Japan}
\affiliation{Institute of Astronomy, Graduate School of Science, The University of Tokyo, 2-21-1 Osawa, Mitaka, Tokyo 181-0015, Japan}

\author{Kai Murai}
\affiliation{Institute for Cosmic Ray Research, The University of Tokyo, 5-1-5 Kashiwanoha, Kashiwa, Chiba 277-8582, Japan}

\author{Moka Nishigaki}
\affiliation{Department of Astronomical Science, SOKENDAI (The Graduate University for Advanced Studies), Osawa 2-21-1, Mitaka, Tokyo, 181-8588, Japan}

\author[0000-0003-3228-7264]{Masato Onodera}
\affiliation{Subaru Telescope, National Astronomical Observatory of Japan, National Institutes of Natural Sciences (NINS), 650 North A'ohoku Place, Hilo, HI 96720, USA}
\affiliation{Department of Astronomical Science, SOKENDAI (The Graduate University for Advanced Studies), Osawa 2-21-1, Mitaka, Tokyo, 181-8588, Japan}

\author[0000-0001-9011-7605]{Yoshiaki Ono}
\affiliation{Institute for Cosmic Ray Research, The University of Tokyo, 5-1-5 Kashiwanoha, Kashiwa, Chiba 277-8582, Japan}

\author{Michael Rauch}
\affiliation{Carnegie Observatories, 813 Santa Barbara Street, Pasadena, CA 91101, USA}

\author{Tomoki Saito}
\affiliation{Nishi-Harima Astronomical Observatory, Centre for Astronomy, University of Hyogo, 407-2 Nishigaichi, Sayo, Sayo-gun, Hyogo 679-5313}

\author{Rin Sasaki}
\affiliation{Department of Physical Science and Materials Engineering, Faculty of Science and Engineering, Iwate University \\
3-18-34 Ueda, Morioka, Iwate 020-8550, Japan}

\author[0000-0002-7043-6112]{Akihiro Suzuki}
\affiliation{Research Center for the Early Universe, The University of Tokyo, 7-3-1 Hongo, Bunkyo, Tokyo 113-0033, Japan}

\author[0000-0001-8416-7673]{Tsutomu T.\ Takeuchi}
\affiliation{Division of Particle and Astrophysical Science, Nagoya University, Furo-cho, Chikusa-ku, Nagoya 464--8602, Japan}
\affiliation{The Research Center for Statistical Machine Learning, the Institute of Statistical Mathematics, 10-3 Midori-cho, Tachikawa, Tokyo 190---8562, Japan}

\author{Hiroya Umeda}
\affiliation{Institute for Cosmic Ray Research, The University of Tokyo, 5-1-5 Kashiwanoha, Kashiwa, Chiba 277-8582, Japan}
\affiliation{Department of Physics, Graduate School of Science, The University of Tokyo, 7-3-1 Hongo, Bunkyo, Tokyo 113-0033, Japan}

\author{Masayuki Umemura}
\affiliation{Center for Computational Sciences, University of Tsukuba, Ten-nodai, 1-1-1 Tsukuba, Ibaraki 305-8577, Japan}

\author{Kuria Watanabe}
\affiliation{Department of Astronomical Science, SOKENDAI (The Graduate University for Advanced Studies), Osawa 2-21-1, Mitaka, Tokyo, 181-8588, Japan}

\author[0000-0001-6229-4858]{Kiyoto Yabe}
\affiliation{Kavli Institute for the Physics and Mathematics of the Universe (WPI), University of Tokyo, Kashiwa, Chiba 277-8583, Japan}

\author{Yechi Zhang}
\affiliation{Institute for Cosmic Ray Research, The University of Tokyo, 5-1-5 Kashiwanoha, Kashiwa, Chiba 277-8582, Japan}
\affiliation{Department of Physics, Graduate School of Science, The University of Tokyo, 7-3-1 Hongo, Bunkyo, Tokyo 113-0033, Japan}




\begin{abstract}
We present kinematics of 6 local extremely metal-poor galaxies (EMPGs) with low metallicities ($0.016-0.098\ Z_{\odot}$) and low stellar masses ($10^{4.7}-10^{7.6} M_{\odot}$).
Taking deep medium-high resolution ($R\sim7500$) integral-field spectra with 8.2-m Subaru, we resolve the small inner velocity gradients and dispersions of the EMPGs with H$\alpha$ emission. 
Carefully masking out sub-structures originated by inflow and/or outflow,
we fit 3-dimensional disk models to the observed H$\alpha$ flux, velocity, and velocity-dispersion maps.
All the EMPGs show rotational velocities ($v_{\rm rot}$) of 5--23 km s$^{-1}$ smaller than the velocity dispersions
($\sigma_{0}$) of 17--31 km s$^{-1}$, indicating dispersion-dominated \tcrc{($v_{\rm rot}/\sigma_{0}=0.29-0.80<1$) systems affected} by inflow and/or outflow.
Except for two EMPGs with large uncertainties,
we find that the EMPGs have very large gas-mass fractions of $f_{\rm gas}\simeq 0.9-1.0$. Comparing our results with other H$\alpha$ kinematics studies, we find that $v_{\rm rot}/\sigma_{0}$ decreases and $f_{\rm gas}$ increases with decreasing metallicity, decreasing stellar mass, and increasing specific star-formation rate.
\tcrb{\tcrc{We also find} that simulated high-$z$ ($z\sim 7$) forming galaxies have gas fractions and dynamics similar to the observed EMPGs. Our EMPG observations and the simulations suggest that primordial galaxies are gas-rich dispersion-dominated systems\tcrc{, which} would be identified by the forthcoming James Webb Space Telescope (JWST) observations at $z\sim 7$.}

\end{abstract}



\section{Introduction} \label{sec:intro}
Primordial galaxies \tcra{characterized by low metallicities and low stellar masses} are important to understand galaxy formation. \tcra{Cosmological and hydrodynamical simulations \citep[e.g.,][]{Wise2012a,Yajima2022} have} predicted that primordial galaxies at \tcra{$z\sim10$} would form in \tcra{low-mass} halos with halo masses of \tcra{$\sim10^{8}\ M_{\odot}$}.
\tcra{Such simulated primordial galaxies} at $z\gtrsim7$ show low \tcra{gas-phase} metallicities of $\lesssim1$\% of the solar abundance ($Z_{\odot}$), specific star-formation rates (sSFRs) of $\sim100$ Gyr$^{-1}$, and low stellar masses of $\lesssim10^{6}\ M_{\odot}$.
\tcra{Despite \tcrc{their scientific relevance}, it is difficult to observe primordial galaxies due to their faintness.}
\citeauthor{Isobe2022} (\citeyear{Isobe2022}; hereafter Paper IV) have estimated an H$\alpha$ flux of a primordial galaxy with $M_{*}=10^{6}\ M_{\odot}$ \tcrd{as a function of} redshift, and demonstrated that even the current-best spectrographs such as Keck/LRIS or Keck/MOSFIRE cannot observe the primordial galaxy at $z\gtrsim0.5$ without any gravitational lensing magnification \citep[e.g.,][]{Kikuchihara2020}.

Kinematics of high-$z$ primordial galaxies \tcrb{can provide us a hint of what kind of mechanism (e.g., inflow/outflow) would impact on the early galaxy formation.}
\tcrc{We are still lacking a good handle on the detailed gas dynamical state, often quantified in broad terms by the relative level of rotation, via the $v_{\rm rot}/\sigma_{0}$ ratio \citep[e.g.,][]{ForsterSchreiber2009}.}
Integral-field \tcra{unit (IFU)} observations \citep[e.g.,][]{Wisnioski2015,Herrera-Camus2022arxiv} have reported that star-forming galaxies show \tcra{$v_{\rm rot}/\sigma_{0}$ ratios} decreasing from $v_{\rm rot}/\sigma_{0}\sim10$ to $\sim2$ with increasing redshift from $z\sim0$ to 5.5\tcra{, while such observations are currently limited to massive ($\sim10^{10}\ M_{\odot}$) galaxies}.
\tcra{It is important to directly determine whether low-mass ($\lesssim10^{6}\ M_{\odot}$)} primordial galaxies are \tcra{truly dominated by dispersion}.

Complementary to the high-$z$ kinematics studies, some studies have reported $v_{\rm rot}/\sigma_{0}$ values of local galaxies with lower stellar masses.
\tcra{Local galaxies are advantageous for conducting deep observations with high spectral and spatial resolutions.}
\tcra{\tcrd{The} SH$\alpha$DE \tcrd{survey} \citep{Barat2020} has made a significant progress} in extracting $v_{\rm rot}$ and $\sigma_{0}$ values of local dwarf galaxies with stellar masses down to $\sim10^{6}\ M_{\odot}$.
\tcrd{These} \tcra{observations suggest} that the ratio $v_{\rm rot}/\sigma_{0}$ decreases \tcra{down to $\lesssim1$} with decreasing $M_{*}$ down to $\sim10^{6}\ M_{\odot}$.
However, the SH$\alpha$DE galaxies have gas-phase oxygen metallicities\footnote{Drawn from the SDSS MPA-JHU catalog \citep{Tremonti2004,Brinchmann2004}.} higher than $12+\log(\rm O/H)\sim7.69$ that correspond to $Z\sim0.1\ Z_{\odot}$ \citep{Asplund2021}, which are still $\gtrsim1$ dex higher than that of the simulated high-$z$ primordial galaxy \citep{Wise2012a}. 

\tcra{To understand the kinematic properties of chemically-primordial galaxies}, we investigate H$\alpha$ kinematics of local galaxies with $Z\leq0.1\ Z_{\odot}$ that are often referred to as extremely metal-poor galaxies \citep[e.g.,][]{Kunth2000,Izotov2012}, abbreviated as EMPGs (\citealt{Kojima2020}, hereafter Paper I).
Although EMPGs become rarer toward lower redshifts \citep{Morales-Luis2011}, various studies have reported the presence of EMPGs in the local universe.
Representative and well-studied EMPGs are SBS0335$-$052 \citep{Izotov2009}, AGC198691 \citep{Hirschauer2016}, Little Cub \citep{Hsyu2017}, DDO68 \citep{Pustilnik2005}, IZw18 \citep{Izotov1998}, and Leo P \citep{Skillman2013}.
\citet{Izotov2018a} have pinpointed J0811+4730 with a low metallicity of 0.02 $Z_{\odot}$. 

Recently, a project ``Extremely Metal-Poor Representatives Explored by the Subaru Survey (EMPRESS)'' \tcra{has been launched (Paper I).
EMPRESS aims to select} faint EMPG photometric candidates from Subaru/Hyper Suprime-Cam (HSC; \citealt{Miyazaki2018}) \tcra{deep optical} ($i_{\rm lim}=26$ mag; \citealt{Aihara2019}) images, which are $\sim2$ dex deeper than those of SDSS.
Conducting follow-up spectroscopic observations of the EMPG photometric candidates, EMPRESS has identified \tcrd{new} 12 EMPGs with low stellar masses of $10^{4.2}$--$10^{6.6}\ M_{\odot}$ (Papers I, IV; \citealt{Nakajima2022arXiv}, hereafter Paper V; \citealt{Xu2022}, hereafter Paper VI).
Remarkably, J1631+4426 \tcra{has been reported to have} a metallicity of 0.016 $Z_{\odot}$, which is the lowest metallicity idenitifed so far (Paper I).

Including the 12 low-mass EMPGs found by EMPRESS, Paper V summarizes 103 local EMPGs identified so far whose metallicities are accurately measured with the direct-temperature method \citep[e.g.,][]{Izotov2006}.
The 103 EMPGs show low metallicities of 0.016--0.1 $Z_{\odot}$, low stellar masses of $\sim10^{4}$--$10^{8}\ M_{\odot}$, and high sSFRs of $\sim1$--$400$ Gyr$^{-1}$ (Paper V).
These features resemble the simulated primordial galaxy at $z\gtrsim7$ \citep{Wise2012a}, suggesting that EMPGs would be good local analogs of high-$z$ primordial galaxies \tcra{(but see also \citealt{Isobe2021}, hereafter Paper III)}.

This paper is the ninth paper of EMPRESS, reporting H$\alpha$ kinematics of EMPGs observed with Subaru/\tcra{Faint Object Camera and Spectrograph (FOCAS)} IFU \tcrb{\citep{Ozaki2020}} in a series of the Subaru Intensive Program entitled EMPRESS 3D (PI: M. Ouchi).
So far, EMPRESS has released 8 papers related to EMPGs, each of which reports the survey design (Paper I), high Fe/O ratios suggestive of massive stars (\citealt{Kojima2021}, hereafter Paper II; Paper IV), morphology (Paper III), low-$Z$ ends of metallicity diagnostics (Paper V), outflows (Paper VI), the shape of incident spectrum that reproduces high-ionization lines (\citealt{Umeda2022}, hereafter Paper VII), and the primordial He abundance (\citealt{Matsumoto2022}, hereafter Paper VIII).

This paper is organized as follows.
Section \ref{sec:sample} explains our observational targets.
The observations are summarized in Section \ref{sec:obs}.
Section \ref{sec:map} reports H$\alpha$ flux, velocity, and velocity-dispersion maps of our targets. 
Our kinematic analysis is described in Section \ref{sec:analysis}.
Section \ref{sec:result} lists our results.
We discuss kinematics of primordial galaxies in Section \ref{sec:dis}.
Our findings are summarized in Section \ref{sec:sum}.
Throughout this paper, magnitudes are in the AB system \citep{Oke1983}.
We assume a standard $\Lambda$CDM cosmology with parameters of ($\Omega_{\rm m}$, $\Omega_{\rm \Lambda}$, $H_{0}$) = (0.3, 0.7, 70 km ${\rm s}^{-1}$ ${\rm Mpc}^{-1}$). 
The solar metallicity $Z_{\odot}$ is defined by $12+\log(\rm O/H)=8.69$ \citep{Asplund2021}.

\section{Sample} \label{sec:sample}
\begin{table*}[t]
    \begin{center}
    \caption{Properties of the observed 6 EMPGs}
    \label{tab:fund}
    \begin{tabular}{ccccccccc} \hline \hline
		\# & Name & R.A. & Decl. & Redshift & $12+\log(\rm O/H)$ & $\log(M_{*})$ & \tcre{$\log(\rm SFR)$} & \tcrf{$\log(\rm sSFR)$} \\
		& & hh:mm:ss & dd:mm:ss & & & $M_{\odot}$ & $M_{\odot}$ yr$^{-1}$ & Gyr$^{-1}$ \\
		(1) & (2) & (3) & (4) & (5) & (6) & (7) & (8) & (9) \\ \hline
		1 & J1631+4426 & 16:31:14.24 & +44:26:04.43 & 0.0313 & $6.90\pm0.03^{\textcolor{blue}{1}}$ & $5.9^{\textcolor{blue}{1}}$ & $-1.3^{\textcolor{blue}{1}}$ & \tcrf{1.8} \\
		2 & IZw18NW & 09:34:02.03 & +55:14:28.07 & 0.0024 & $7.16\pm0.01^{\textcolor{blue}{2}}$ & $7.1^{\textcolor{blue}{3}}$ & $-1.4^{\textcolor{blue}{4}}$ & \tcrf{0.5} \\
		3 & SBS0335$-$052E & 03:37:44.06 & $-$05:02:40.19 & 0.0135 & $7.22\pm0.07^{\textcolor{blue}{5}}$ & $7.6^{\textcolor{blue}{5}}$ & $-0.4^{\textcolor{blue}{4}}$ & \tcrf{1.0} \\
		4 & HS0822+3542 & 08:25:55.44 & +35:32:31.92 & 0.0020 & $7.45\pm0.02^{\textcolor{blue}{6}}$ & \tcre{4.6}$^{\textcolor{blue}{7}}$ & $-2.2^{\textcolor{blue}{8}}$ & \tcrf{2.2} \\
		5 & J1044+0353 & 10:44:57.79 & +03:53:13.15 & 0.0130 & $7.48\pm0.01^{\textcolor{blue}{6}}$ & \tcre{6.0}$^{\textcolor{blue}{7}}$ & $-0.9^{\textcolor{blue}{9}}$ & \tcrf{2.1} \\ 
		6 & J2115$-$1734 & 21:15:58.33 & $-$17:34:45.09 & 0.0230 & $7.68\pm0.01^{\textcolor{blue}{1}}$ & $6.6^{\textcolor{blue}{1}}$ & $0.3^{\textcolor{blue}{1}}$ & \tcrf{2.7} \\ \hline
    \end{tabular}
    \end{center}
    \tablecomments{(1) Number. (2) Name. (3) Right ascension in J2000. (4) \tcra{Declination} in J2000. (5) Redshift. (6) Metallicity. (7) Stellar mass. \tcre{(8) Star-formation rate based on H$\alpha$ luminosity.} (9) Specific star-formation rate. References: $^{\textcolor{blue}{1}}$Paper I; $^{\textcolor{blue}{2}}$\citet{Izotov1998}; $^{\textcolor{blue}{3}}$\citet{Annibali2013}; 
    $^{\textcolor{blue}{4}}$\citet{Thuan1997}; 
    $^{\textcolor{blue}{5}}$\citet{Izotov2009}; 
    $^{\textcolor{blue}{6}}$\citet{Kniazev2003}; 
    $^{\textcolor{blue}{7}}$\textbf{This paper}; 
    $^{\textcolor{blue}{8}}$\citet{Kniazev2000}; 
    $^{\textcolor{blue}{9}}$\citet{Berg2016}}
\end{table*}

\tcra{We select 6 EMPGs visible on the observing nights (Section \ref{subsec:obs}) and} having relatively strong H$\beta$ fluxes at a given metallicity so that we obtain kinematics of the EMPGs with high signal-to-noise (SN) ratios.
\tcrb{Details of the observational strategy will be reported in Xu et al. (in prep.).}
The 6 EMPGs consists of J1631+4426 (Paper I), IZw18 \citep[e.g.,][]{Searle1972}, SBS0335$-$052E \citep[e.g.,][]{Izotov1997}, HS0822+3542 \citep{Kniazev2000}, J1044+0353 \citep{Papaderos2008}, and J2115$-$1734 (Paper I).

We note that the 4 EMPGs other than J1631+4426 or J2115$-$1734 have radio and/or optical integral-field spectroscopy conducted by previous studies.
IZw18 has H\,{\sc i} observations with VLA \citep{Lelli2012}.
SBS0335$-$052E also has VLA H\,{\sc i} observations \citep{Pustilnik2001} as well as VLT/MUSE H$\alpha$ observations with \tcrc{a} spectral resolution of \tcra{$R\sim3000$} \citep{Herenz2017}. 
Our observations with FOCAS IFU are complementary to those previous observations of IZw18 and SBS0335$-$052E \tcrb{because of our higher spectral resolution ($R\sim7500$, Section \ref{subsec:obs})}.
HS0822+3542 and J1044+0353 have H$\alpha$ observations with 6-m BTA/Fabry-Perot interferometer having $R\sim8000$ \citep{Moiseev2010}. 
FOCAS IFU can detect emission lines $\sim2$ times fainter than those of Fabry-Perot interferometer with a similar spectral resolution.

Properties of the observed 6 EMPGs are listed in Table \ref{tab:fund}.
The 6 EMPGs have low metallicities of $12+\log(\rm O/H)=6.90$--7.68, low stellar masses of $\log(M_{*}/M_{\odot})=4.7$--7.6, and high specific star-formation rates of \tcrf{$\log(\rm sSFR/Gyr^{-1})=0.5$--2.7}.
These properties indicate that the 6 EMPGs are \tcrc{the best available} analogs of primordial galaxies in the early universe. 
We emphasize that the 6 EMPGs include J1631+4426 having $12+\log(\rm O/H)=6.90$ (Paper I), which is the lowest gas-phase metallicity among galaxies identified so far.

\section{Observations and \tcra{data analysis}} \label{sec:obs}
\subsection{\tcra{Observations} \tcrb{and Data Reduction}} \label{subsec:obs}
This section reports our spectroscopic observations with FOCAS IFU \tcrb{\citep{Ozaki2020}}.
FOCAS IFU is an IFU with an image slicer installed in FOCAS \citep{Kashikawa2002} mounted on a Cassegrain focus of the Subaru \tcra{8.2}-m telescope.
The large diameter of the Subaru Telescope allows FOCAS IFU to perform deep integral-field spectroscopy with a 5$\sigma$ limiting flux of $\sim1\times10^{-17}$ erg s$^{-1}$ cm$^{-2}$ arcsec$^{-2}$ \citep{Ozaki2020}\footnote{\tcra{Under the assumptions of an 1-hour exposure and an extended source whose intrinsic line width is negligible with respect to the instrumental broadening.}}.
\tcra{The} FoV of FOCAS IFU is \tcra{$13.\!\!\arcsec5$} (slice length direction; hereafter X direction) $\times$ $10.\!\!\arcsec0$ (slice width direction; hereafter Y direction).
\tcrb{The} pixel scale in a reduced data cube is $0.\!\!\arcsec215$ (X direction) and $0.\!\!\arcsec435$ (Y direction).

We carried out spectroscopy with FOCAS IFU for the 6 EMPGs (Section \ref{sec:sample}).
\tcra{The observing nights were} 2021 August 13, November 24, and December 13.
We set pointing positions so that the whole structures of EMPGs are covered with single FoVs.

We used the mid-high-dispersion grism of VPH680 \tcrb{offering} the spectral resolution of $R\sim7500$.
We took comparison frames of the ThAr lamp.
We observed Feige67, HZ44, BD28, Feige110, and Feige34 as standard stars.
All the nights were clear with seeing sizes of $\sim0.\!\!\arcsec7$.

We use a reduction pipeline software of FOCAS IFU\footnote{\url{https://www2.nao.ac.jp/~shinobuozaki/focasifu/}} based on PyRAF \citep{Tody1986} and Astropy \citep{Astropy2013}.
The software performs bias subtraction, flat-fielding, cosmic-ray cleaning, sky subtraction, wavelength calibration \tcre{with the comparison frames}, and flux calibration.
We estimate flux errors containing read-out noises and photon noises of sky and object emissions.

It should be noted that there remain systematic velocity differences among the slices even after the wavelength calibration with the comparison frames.
We conducted additional wavelength calibration with sky emission lines around observed wavelengths of H$\alpha$.

\subsection{\tcrf{Flux, Velocity, and \tcrg{Velocity} Dispersion Measurement}} \label{subsec:meas}
\begin{figure}[t]
    \centering
    \includegraphics[width=8.0cm]{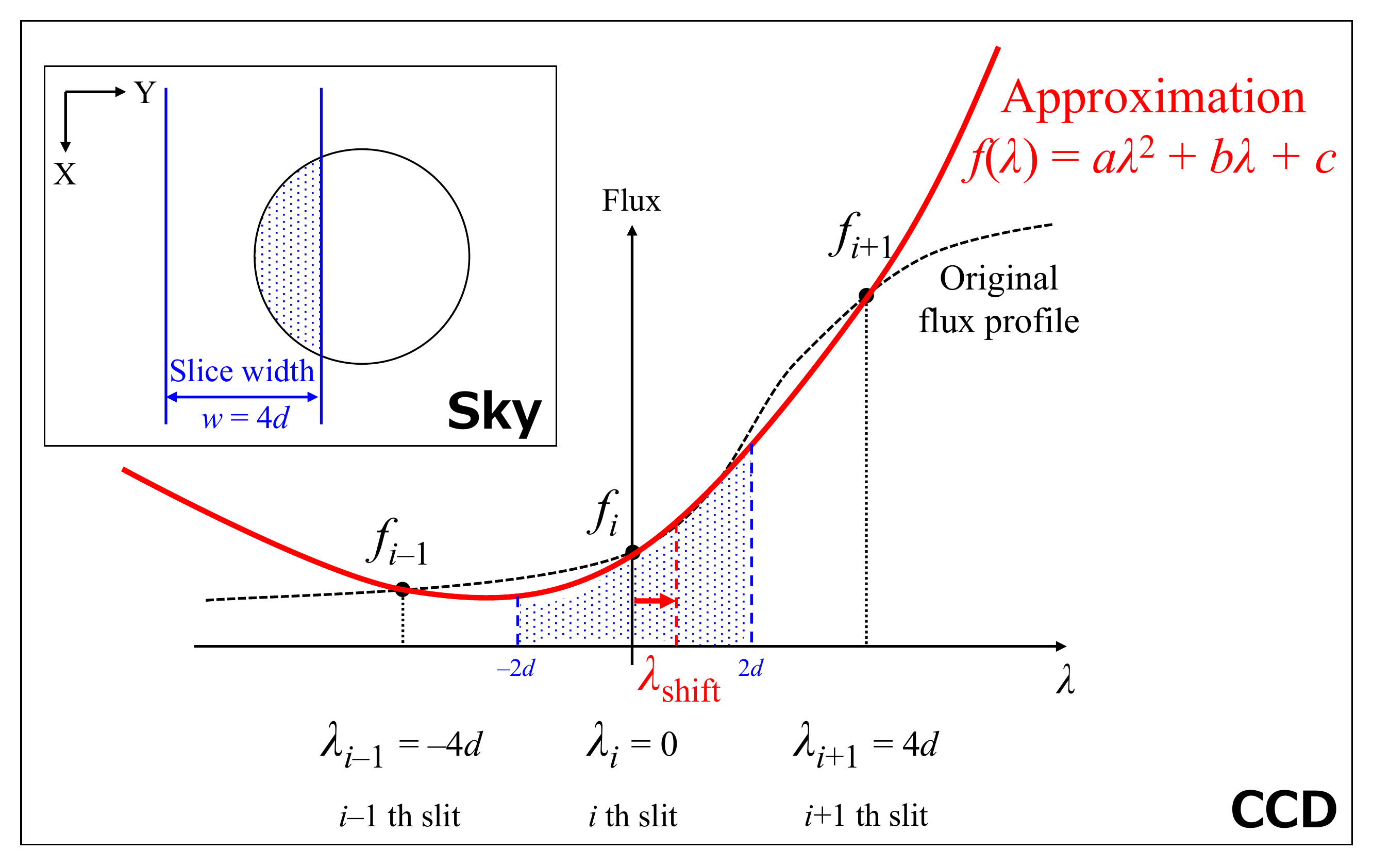}
    \caption{Explanation of the slit-width effect. \tcra{The inset panel shows a schematic figure of an object on the celestial sphere. The X (Y) direction denotes the slice length (width) direction. The \tcrb{blue solid} lines describe the position of the single slice. The light coming into the slice (blue shaded) is projected onto the CCD, where the Y direction in the sky is parallel to the wavelength ($\lambda$) direction on the CCD. The original spatial flux profile projected onto the CCD (black dashed curve) provides fluxes of ($i-1$), $i$, and ($i+1$)th slices ($f_{i-1},\ f_{i}$, and $f_{i+1}$, respectively). The original flux profile is assumed to be approximated by the quadratic function $f(\lambda)=a\lambda^{2}+b\lambda+c$ (red solid curve) within the $\lambda$ range from $-2d$ to $2d$. We estimate the systematic wavelength shift originating from the slit-width effect ($\lambda_{\rm shift}$) from a barycenter of the flux following $f(\lambda)$ intercepted by the slice.}}
    \label{fig:swe}
\end{figure}

\begin{figure*}[t]
    \centering
    \includegraphics[width=18.0cm]{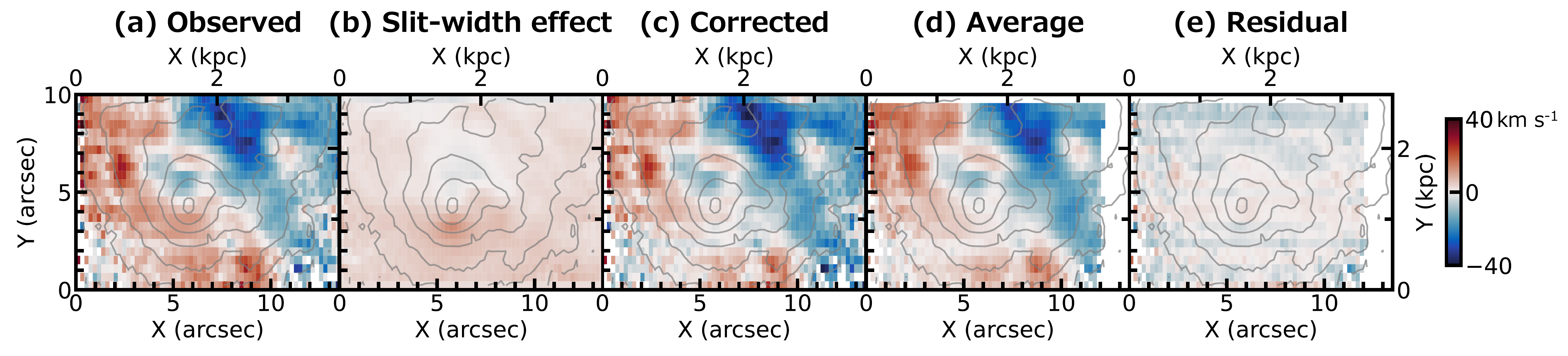}
    \caption{Confirmation of the slit-width effect correction using SBS0335$-$052E. \tcre{\tcrf{(a)} Observed velocity map. \tcrf{(b)} Velocity shift generated by the slit-width effect. \tcrf{(c)}} Velocity map corrected for the slit-width effect. \tcrf{(d)} Average of the 2 velocity maps whose position angles are different by 180 degree. \tcrf{(e)} Residual of the corrected map and the average map. The gray contours illustrate SN ratios in the order of 5, 10, 20, ... from the outside.}
    \label{fig:swecorr}
\end{figure*}

\begin{figure}[t]
    \centering
    \includegraphics[width=8.0cm]{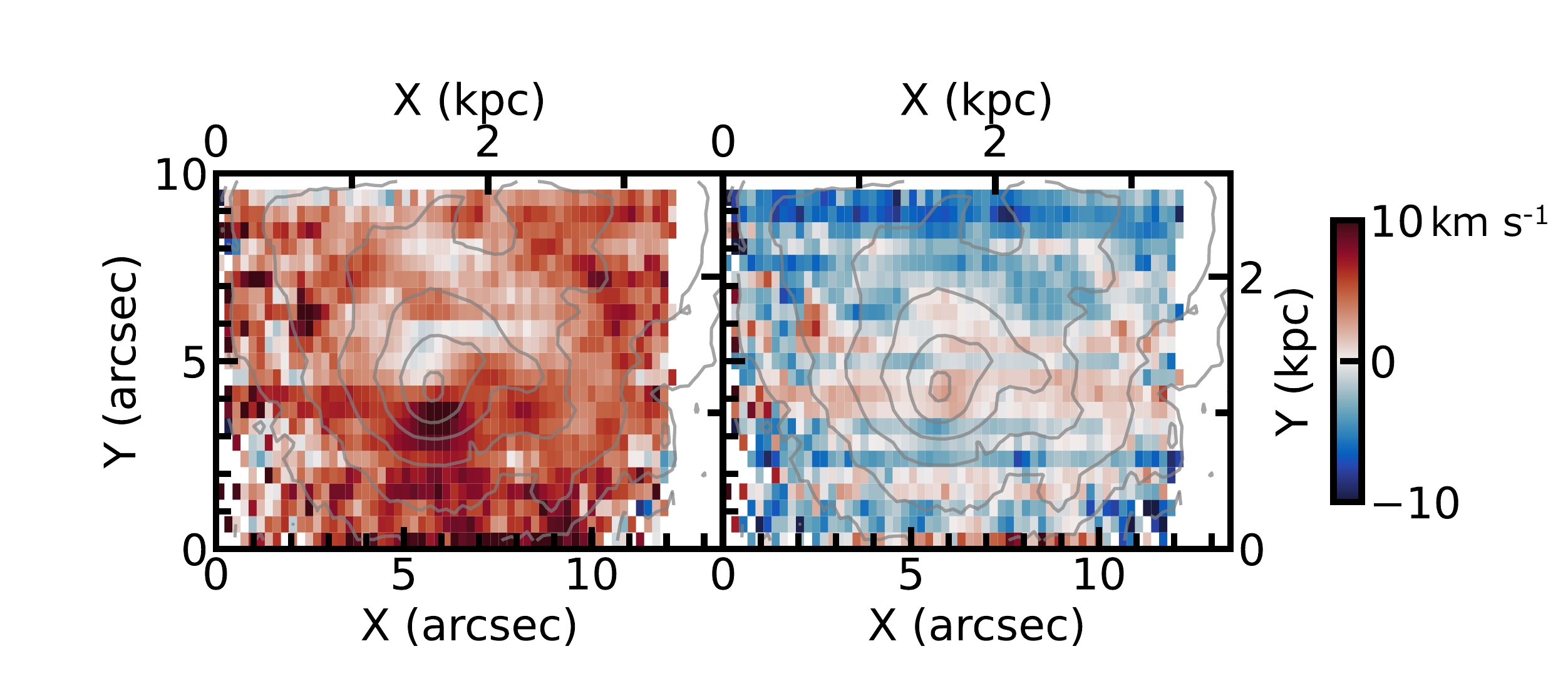}
    \caption{\tcre{(Left) Residual of the observed velocity map and the average map in Figure \ref{fig:swecorr}. (Right) Residual of the corrected map and the average map, i.e., same as \tcrf{panel (e)} in Figure \ref{fig:swecorr}, but with the smaller velocity range.}}
    \label{fig:sweresc}
\end{figure}

At all the spaxels, we fit H$\alpha$ lines using \tcrc{a} single Gaussian function (+ constant) to measure H$\alpha$ fluxes, (relative) velocities, and velocity dispersions.
We derive \tcrc{a} line-spread function (LSF) by measuring widths of sky lines.
\tcra{A typical sky line has a full-width half maximum (FWHM) of $\sim0.8$--1 \AA.
Assuming that both the H$\alpha$ lines from the EMPGs and the sky lines agree well with the Gaussian function,} we obtain \tcra{the} intrinsic velocity dispersions by subtracting the LSFs from the observed velocity dispersions quadratically.
\tcra{We confirm that the assumption is reasonable for most of the H$\alpha$ lines and the sky lines other than some turbulent regions with multiple velocity components (Section \ref{sec:map}).}

\tcrf{We estimate the errors of the velocities and velocity dispersions by running the Monte Carlo simulations similar to the procedure of \citet{Herenz2016}.
We make 100 mock data cubes from our data cubes perturbed by the noise data cubes, and measure the velocities and velocity dispersions from each mock data cube.
We regard the standard deviations of the derived velocities and velocity dispersions as the errors.
The errors include the uncertainties of the LSFs, the additional wavelength calibration with the sky lines (Section \ref{subsec:obs}), and the slit-width effect correction (see Section \ref{subsec:swe}), because we perform these corrections using each mock data cube.
\tcrg{Typical uncertainties of the velocity and velocity dispersion in each spaxel are $\sim1.7$ and $\sim1.6$ km s$^{-1}$, respectively.}
We self-consistently obtain the errors of the kinematic parameters in Sections \ref{subsec:shear} and \ref{subsec:galpak} based on the 100 mock data cubes, in the same manner as we measure the errors of the velocities and velocity dispersions.}

\subsection{\tcra{Slit-width Effect Correction}} \label{subsec:swe}
It should be noted that the observed velocities suffer from systematic wavelength shifts known as the slit-width effect \citep[e.g.,][]{Bacon1995}.
Figure \ref{fig:swe} illustrates the mechanism of the slit-width effect.
The artificial wavelength shift is caused by a flux gradient in the Y direction \tcrc{(in each slice) that follows the  dispersion (wavelength) axis ($\lambda$ direction thereafter) on the CCD.}

Assuming a 2D Gaussian function for the (spatial) flux profile (i.e., assuming a point source), \citet{Bacon1995} have derived the wavelength shifts originating from the slit-width effect.
\tcra{Considering that all our targets are extended and complex sources, we expanded the method to \tcrc{more flexible} flux profiles.}
The wavelength shift $\lambda_{\rm shift}$ can be estimated from a barycenter of the flux intercepted by the slice as follows:
\begin{equation}
    \label{equ:swe}
    \lambda_{\rm shift}=\frac{\int_{-0.5w}^{0.5w}\lambda f(\lambda){\rm d}\lambda}{\int_{-0.5w}^{0.5w}f(\lambda){\rm d}\lambda},
\end{equation}
where $w$ and $f(\lambda)$ are a slice width projected onto the CCD and a flux profile in the Y direction, respectively.
Because the pseudo slit width is sampled by 4 pixels for each slice, $w$ in the unit of \AA\ is given by $w=4d$, where $d$ is a dispersion in the unit of \AA\ pixel$^{-1}$.

We approximate $f(\lambda)$ by a quadratic function $f(\lambda)=a\lambda^{2}+b\lambda+c$ \tcra{so that the function form is determined by the 3 points ($\lambda_{i-1}$, $f_{i-1}$), ($\lambda_{i}$, $f_{i}$), and ($\lambda_{i+1}$, $f_{i+1}$) in Figure \ref{fig:swe}, where $f_{i-1}$, $f_{i}$, and $f_{i+1}$ represent fluxes of $(i-1)$th, $i$th, and $(i+1)$th slices, respectively}.
In this case, $\lambda_{\rm shift}$ can be calculated as follows:
\begin{eqnarray}
    \label{equ:swe2}
    \lambda_{\rm shift}&=&\frac{\int_{-2d}^{2d}\lambda (a\lambda^{2}+b\lambda+c){\rm d}\lambda}{\int_{-2d}^{2d}(a\lambda^{2}+b\lambda+c){\rm d}\lambda} \nonumber \\
    &=&\frac{4bd^{2}}{4ad^{2}+3c}.
\end{eqnarray}
We derive $a$, $b$, and $c$ from the equations below:
\begin{eqnarray}
    \label{equ:const}
    a&=&\frac{f_{i-1}-2f_{i}+f_{i+1}}{16d^{2}} \nonumber \\
    b&=&\frac{f_{i+1}-f_{i-1}}{4d} \\
    c&=&f_{i}. \nonumber
\end{eqnarray}
Using Equations \ref{equ:swe2} and \ref{equ:const}, we obtain
\begin{equation}
    \label{equ:swe3}
    \lambda_{\rm shift}=\frac{4(f_{i+1}-f_{i-1})}{f_{i-1}+10f_{i}+f_{i+1}}d.
\end{equation}
We infer from Equation \ref{equ:swe3} that high-dispersion dispersers make the slit-width effect weak.
VPH680 has only $d=0.22$ \AA\ pixel$^{-1}$, which results in velocity shifts produced by the slit-width effect of only $\sim\pm10$ km s$^{-1}$ under the assumption of the seeing size of $0.\!\!\arcsec7$.
\tcra{Although the velocity shift ($\sim10$ km s$^{-1}$) is small, it is important to correct the slit-width effect because the low-mass EMPGs are expected to have small velocity gradients (Section \ref{sec:map}).}

\tcre{\tcrf{Panel (a)} of Figure \ref{fig:swecorr} shows an observed velocity map of one of the EMPGs (SBS0335$-$052E), and \tcrf{panel (b)} represents velocity shifts caused by the slit-width effect.
A velocity map corrected for the slit-width effect is shown in \tcrf{panel (c)}.}

To test our slit-width effect correction, we observed SBS0335$-$052E with 2 frames whose position angles are different by 180 degree.
We expect that an average of the 2 velocity maps obtained from the 2 frames cancel out the slit-width effect.
\tcrf{Panel (d)} of Figure \ref{fig:swecorr} shows the average map, and \tcrf{panel (e)} represents residuals of the corrected map and the average map.
\tcre{The residuals are smaller than the velocity shifts caused by the slit-width effect.
Figure \ref{fig:sweresc} compares residuals of the observed velocity map and the average map (i.e., residuals NOT corrected for the slit-width effect; left) and residuals of the corrected velocity map and the average map (i.e., residuals corrected for the slit-width effect; right).
Figure \ref{fig:sweresc} shows that the residuals get much smaller after the slit-width effect correction.}
The small residuals of $\sim\pm5$ km s$^{-1}$ also indicate that the corrected map agrees well with the average map.
Thus, we conclude that our slit-width correction works well.

\tcre{We evaluate the uncertainty of the slit-width effect correction by performing Monte Carlo simulation based on flux errors.
A typical value of the uncertainty is 0.3 km s$^{-1}$.}

\section{Flux, Velocity, \& Dispersion Maps} \label{sec:map}
\begin{figure*}[t]
    \centering
    \includegraphics[width=18.0cm]{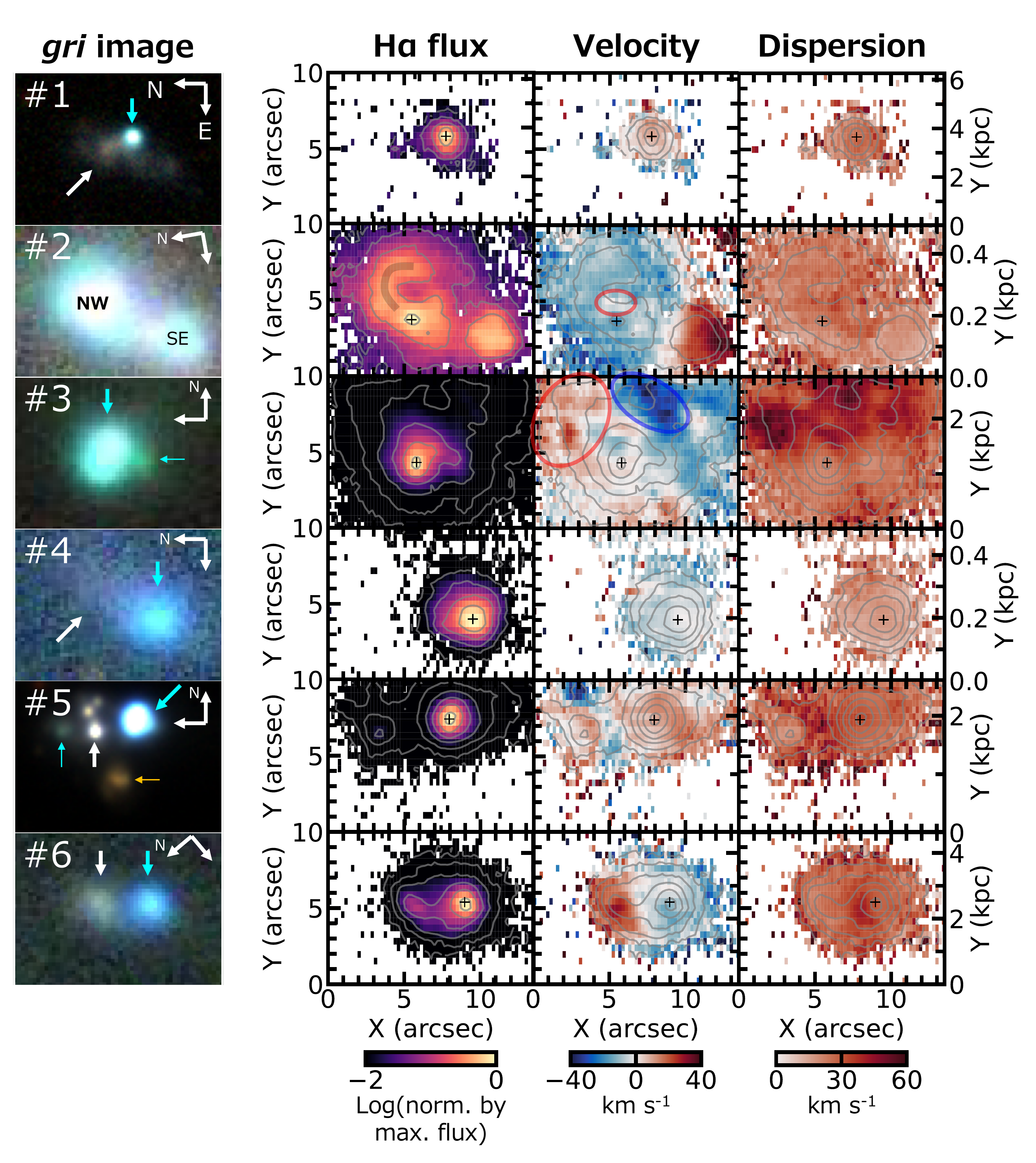}
    \caption{\tcra{(Left) $gri$ images of the 6 EMPGs cut out of the FoV of FOCAS IFU. The numbers correspond to those in Table \ref{tab:fund}. The images of \#1 and 5 are drawn from the HSC-SSP PDR3 \citep{Aihara2022}, while the others are taken from the Pan-STARRS catalog \citep{Flewelling2020}. (Middle left) Observed H$\alpha$ flux maps. The flux values in each flux map are normalized by the maximum flux value of each EMPG. The black crosses show flux peaks. The gray contours are the same as those in Figure \ref{fig:swecorr}. (Middle right) Velocity maps showing relative velocities that fit well in the velocity range from $-40$ to $40$ km s$^{-1}$. The velocity values are corrected for the slit-width effect (Section \ref{subsec:swe}). (Right) Intrinsic velocity-dispersion maps (Section \ref{subsec:swe}).}}
    \label{fig:map}
\end{figure*}

Figure \ref{fig:map} summarizes H$\alpha$ flux, velocity, and velocity dispersion maps of the 6 EMPGs \tcra{as well as $gri$ images cut out of the FoV of FOCAS IFU}.
\tcra{We note that only 2 EMPGs of \#1 and 5 (i.e., J1631+4426 and J1044+0353) have images of the HSC-Subaru Strategic Program (SSP) Public Data Release (PDR) 3 \citep{Aihara2022}.
The deep HSC images of the 2 EMPGs are shown in Figure \ref{fig:map}, while the other $gri$ images are taken from the Pan-STARRS catalog \citep{Flewelling2020}.}
We report \tcra{morphological and} kinematic features of each EMPG below\tcra{, checking the consistency with previous IFU studies}.

{\it \#1 J1631+4426}:
\tcra{The HSC $gri$ image illustrates that} J1631+4426 consists of \tcra{a blue clump \tcre{(indicated by the cyan arrow)} and a white diffuse structure elongated from north to south \tcre{(white arrow)}}.
\tcra{\tcrc{We refer to the white structure as the ``EMPG tail''} (Paper III).
The H$\alpha$ flux map shows that the H$\alpha$ flux of J1631+4426 is dominated by the blue clump, which indicates that star formation in J1631+4426 should mainly occur in \tcrc{that region}.
Paper I has confirmed that the blue clump has the very low metallicity of $12+\log(\rm O/H)=6.90$ (Section \ref{sec:sample}) based on the direct-temperature method, while the EMPG tail can have a metallicity even lower than the blue clump based on the [{\sc O\,iii}]$\lambda$5007/H$\alpha$ ratio \citep{Kashiwagi2021}.}
The velocity map shows that the blue clump is discontinuously redshifted by $\sim20$ km s$^{-1}$ with respect to the EMPG tail, which indicates that the blue clump and the EMPG tail are not in the same kinematic structure.
The blue clump itself shows a weak velocity gradient of $\sim10$ km s$^{-1}$ from east to west, while the velocity dispersion is relatively high ($\sim25$ km s$^{-1}$).

{\it \#2 IZw18}:
IZw18 has 2 main blue clumps of IZw18 \tcrd{Northwest} (NW) and IZw18 \tcrd{Southeast} (SE)\tcra{, both of which have been confirmed to be EMPGs \citep{Izotov1998}}.
IZw18NW is blueshifted by $\sim40$ km s$^{-1}$ with respect to IZw18SE (at around their flux peaks), \tcrc{consistent with the} H\,{\sc i} gas kinematics \citep{Lelli2012}.
We thus think that IZw18NW and IZw18SE are not in the same kinematic structure.
IZw18NW shows a complex morphokinematic structure.
\tcre{In the H$\alpha$ flux map,} we find that the arc-like structure \tcre{indicated by the gray curve} (H$\alpha$ arc; \citealt{Dufour1990}) has a velocity similar to that of IZw18NW, which indicates that the H$\alpha$ arc and IZw18NW belong to the same kinematic structure.
\tcre{In the velocity map,} we also identify a velocity structure that is redshifted by $\sim20$ km s$^{-1}$ with respect to the flux peak of IZw18NW \tcre{(red circle)}.
The velocity structure has a relatively high velocity dispersion of $\sim30$ km s$^{-1}$, which implies that the structure is not \tcra{settled}.
IZw18NW itself does not show a clear bulk rotation \tcra{(i.e., not likely to be dominated by rotation)}.

{\it \#3 SBS0335$-$052E}:
\tcra{The $gri$ image shows that SBS0335$-$052E consists of a large blue clump \tcre{(thick cyan arrow)} and a western subclump \tcre{(thin cyan arrow)}.}
\tcre{In the velocity map,} we confirm that SBS0335$-$052E has a redshifted ($\sim+20$ km s$^{-1}$) region at northeast \tcre{(red circle)} and a blueshifted ($\sim-30$ km s$^{-1}$) region at northwest \tcre{(blue circle)}.
\tcre{The redshifted and blueshifted regions seem connected to the northeast and north filaments} identified with the wide-field MUSE H$\alpha$ observations \citep{Herenz2017}\tcre{, respectively}.
Both regions have relatively high velocity dispersions of $\sim60$ km s$^{-1}$, which agree with the scenario that the 2 structures are created by outflows \citep{Herenz2017}.
The southern area of SBS0335$-$052E generally shows a relatively low velocity dispersion of $\sim30$ km s$^{-1}$, which indicates that the southern area is \tcra{relatively settled}.
The southern area shows a bulk velocity gradient of $\sim20$ km s$^{-1}$ from northwest to southeast.

{\it \#4 HS0822+3542}:
\tcra{HS0822+3542 consists of a blue clump \tcre{(cyan arrow)} and a very diffuse EMPG tail elongated from southeast to northwest \tcre{(white arrow)}.
The H$\alpha$ map indicates that the major star formation occurs in the blue clump.}
We find a very weak velocity gradient of $\sim5$ km s$^{-1}$ \tcra{in the blue clump} from north to south.
HS0822+3542 shows a \tcra{high} velocity dispersion of $\sim20$ km s$^{-1}$, which is comparable to the previous observations \citep{Moiseev2010}.

{\it \#5 J1044+0353}:
\tcra{The deep HSC image shows that J1044+0353 is composed of a western giant blue clump \tcre{(thick cyan arrow)}, an eastern small blue clump \tcre{(thin cyan arrow)}, and 3 white clumps between the 2 blue clumps \tcre{(white arrow)}\footnote{\tcra{Using the 300B grism with a wide wavelength coverage, we detect [{\sc O\,iii}]$\lambda\lambda$4959,5007 lines at $z=0.27$ from the southern red object \tcre{(orange arrow)}. This means that the red object is a background galaxy and thus not related to J1044+0353.}}.
We regard the 3 white clumps as the EMPG tail.
The H$\alpha$ map indicates that the major star formation occurs in the giant blue clump.}
We obtain similar velocity and velocity dispersion maps to \citet{Moiseev2010}'s result\tcra{, while we show more clearly that the small blue clump emits H$\alpha$.}
\tcra{We find that the 2 blue clumps are redshifted by $\sim20$ km s$^{-1}$ with respect to the EMPG tail, which indicate that the 2 blue clumps \tcrb{and the EMPG tail are not in the same kinematic structure}.}
We find that \tcra{the giant blue clump} has a weak velocity gradient of $\sim20$ km s$^{-1}$ from southeast to northwest\tcra{, while the velocity dispersion ($\sim30$ km s$^{-1}$) is quite high.}

{\it \#6 J2115$-$1734}:
J2115$-$1734 consists of a blue clump \tcre{(cyan arrow)} and a \tcra{north-east} EMPG tail \tcre{(white arrow)}.
\tcra{The \tcrd{majority of} star formation is likely to occur in the blue clump.}
The EMPG tail is redshifted by $\sim40$ km s$^{-1}$ with respect to the blue clump.
Although the observed velocity field continuously changes between the blue clump and the EMPG tail, J2115$-$1734 shows a relatively high velocity dispersion of $\sim40$ km s$^{-1}$ \tcrd{originating} from 2 different velocity components.
The blue clump itself has a weak velocity gradient of $\sim20$ km s$^{-1}$ from northeast to southwest with a velocity dispersion of $\sim30$ km s$^{-1}$.

\tcra{In summary, all 6 EMPGs look irregular and dominated by dispersion rather than rotation. \tcrc{Their rotation \tcrd{velocities} are not likely to be significant.} In Section \ref{sec:analysis}, we analyze \tcrd{the} kinematics of the EMPGs more quantitatively.}

\section{Kinematic Analysis} \label{sec:analysis}
\subsection{\tcra{Masking}} \label{subsec:mask}
\begin{figure}[t]
    \centering
    \includegraphics[width=8.0cm]{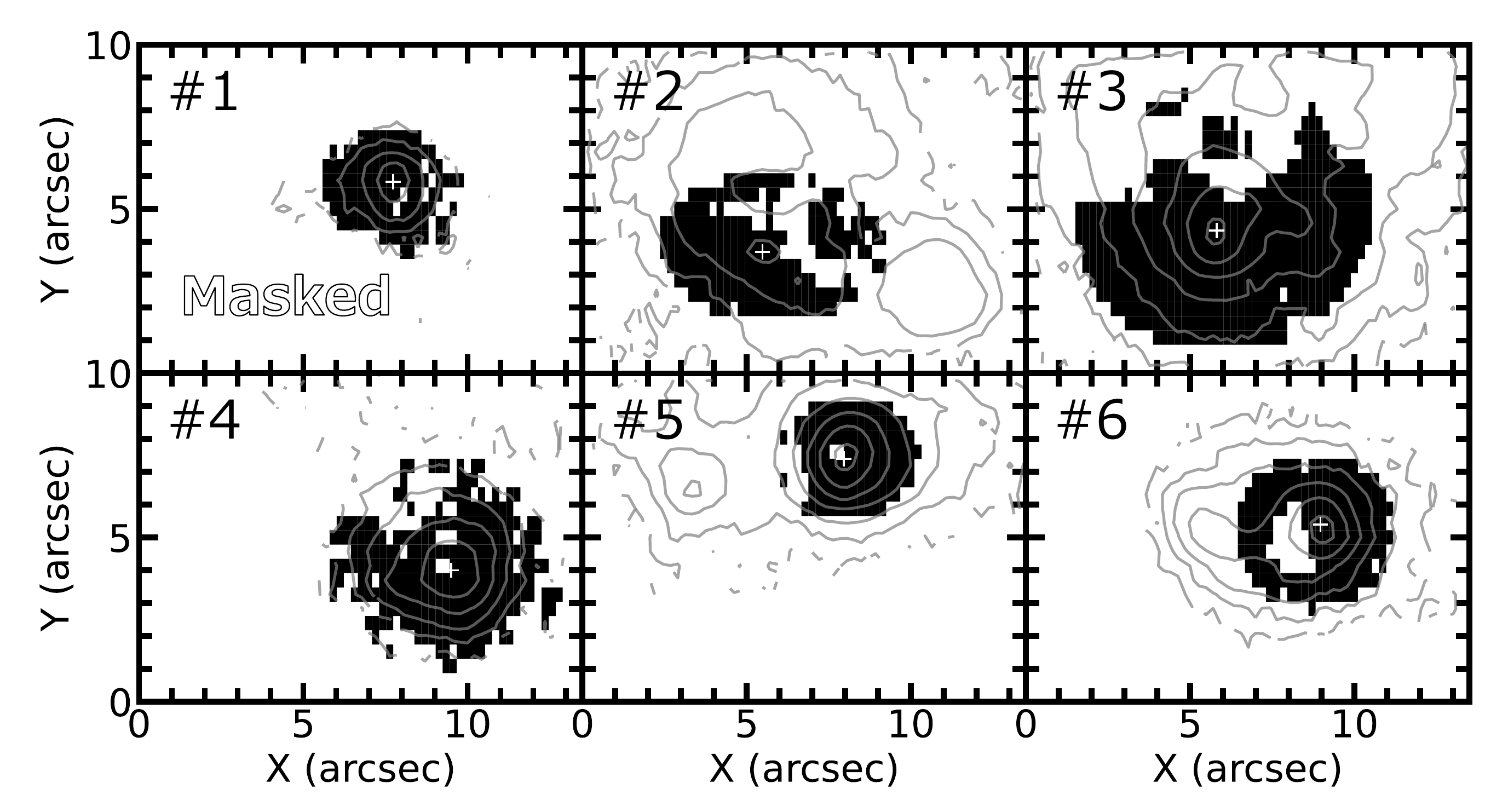}
    \caption{\tcra{Mask map. The white regions correspond to the masked regions (Section \ref{subsec:mask}).}}
    \label{fig:mask}
\end{figure}

\tcra{
\tcrc{Below, we quantify the level of rotation of the EMPGs by assuming that those systems are represented by a single rotating disk.}
The dynamical center of the disk is thought to be located at the H$\alpha$ flux peak of the main blue clump (IZw18NW for IZw18) under the assumptions of the gas-mass dominance on the galactic scale \citep[cf. e.g.,][]{Herrera-Camus2022arxiv} and the empirical positive correlation between the gas mass surface density $\Sigma_{\rm gas}$ and the SFR surface density $\Sigma_{\rm SFR}$ (a.k.a. Kennicutt-Schmidt law; \citealt{Kennicutt1998b}).
We confirm that these assumptions are reasonable by deriving the mass profile of each EMPG (Section \ref{subsec:fgas}).
In the following analysis, we use only the region within the Kron radius of the main blue clump to remove the contamination from EMPG tails (IZw18SE for IZw18).
We also mask spaxels with velocity dispersions higher than the flux-weighted 84th percentile of the distribution of the velocity dispersions because such regions are thought to be highly turbulent \citep{Egorov2021}.
The white regions of Figure \ref{fig:mask} show the masked regions.}

\subsection{\tcra{Non-parametric Method}} \label{subsec:shear}
\begin{table*}[t]
    \begin{center}
    \caption{\tcrf{Kinematic properties of the 6 EMPGs}}
    \label{tab:kin}
    \begin{tabular}{cccccccc} \hline \hline
		\# & Name & $v_{\rm shear}$ & $\sigma_{\rm med}$ & $v_{\rm shear}/\sigma_{\rm med}$ & $v_{\rm rot}$ & $\sigma_{0}$ & $v_{\rm rot}/\sigma_{0}$ \\
		& & km s$^{-1}$ & km s$^{-1}$ & & km s$^{-1}$ & km s$^{-1}$ & \\
		(1) & (2) & (3) & (4) & (5) & (6) & (7) & (8) \\ \hline
		1 & J1631+4426 & $10.3\pm1.3$ & $25.6\pm0.4$ & $0.40\pm0.05$ & $7.9\pm1.8$ & $25.6\pm0.3$ & $0.31\pm0.08$ \\
		2 & IZw18NW & $7.7\pm0.4$ & $23.2\pm0.5$ & $0.33\pm0.02$ & $6.6\pm2.9$ & $22.9\pm0.4$ & $0.29\pm0.13$ \\
		3 & SBS0335$-$052E & $14.3\pm0.4$ & $24.6\pm0.3$ & $0.58\pm0.02$ & $19.7\pm2.9$ & $27.1\pm0.3$ & $0.73\pm0.12$ \\
		4 & HS0822+3542 & $5.6\pm0.7$ & $16.9\pm0.6$ & $0.34\pm0.07$ & $4.5\pm2.9$ & $16.6\pm0.5$ & $0.27\pm0.17$ \\
		5 & J1044+0353 & $5.5\pm0.2$ & $30.8\pm0.3$ & $0.18\pm0.01$ & $14.8\pm4.2$ & $31.4\pm0.3$ & $0.47\pm0.14$ \\ 
		6 & J2115$-$1734 & $9.8\pm0.4$ & $29.7\pm0.2$ & $0.33\pm0.01$ & $23.4\pm8.4$ & $29.3\pm0.1$ & $0.80\pm0.30$ \\ \hline
    \end{tabular}
    \end{center}
    \tablecomments{(1) Number. (2) Name. \tcra{(3) Shearing velocity. (4) Median velocity dispersion. (5) $v_{\rm shear}/\sigma_{\rm med}$.} (6) Rotation velocity. (7) \tcra{Intrinsic} velocity dispersion. (8) $v_{\rm rot}/\sigma_{0}$. }
\end{table*}

\tcra{We calculate a shearing velocity $v_{\rm shear}$ of each EMPG, which is a non-paramteric kinematic property widely used in the literature \citep[e.g.,][]{Law2009,Herenz2016}.
We derive $v_{\rm shear}$ from
\begin{equation}
    \label{equ:shear}
    v_{\rm shear}=\frac{1}{2}(v_{\rm max}-v_{\rm min}),
\end{equation}
where $v_{\rm max}$ ($v_{\rm min}$) is the 95th (5th) percentile of the velocity distribution \citep{Herenz2016}.
We note that high $v_{\rm shear}$ values do not necessarily imply the existence of rotation \tcre{because different velocity components in the complex velocity fields can mimic rotation patterns at small scales.
In this sense, $v_{\rm shear}$ can be regarded as an upper limit of the rotation velocity.}
The global velocity dispersion can be quantified by the flux-weighted median of the distribution of the velocity dispersions ($\sigma_{\rm med}$).
\tcrf{We calculate the errors of the $v_{\rm shear}$ and $\sigma_{\rm med}$ values ($\Delta_{v_{\rm shear}}$ and $\Delta_{\sigma_{\rm med}}$, respectively) by running the Monte Carlo simulations explained in Section \ref{subsec:meas}.}
We list $v_{\rm shear}$ and $\sigma_{\rm med}$ of the 6 EMPGs in Table \ref{tab:kin}.}
\tcrf{The medians of our $v_{\rm shear}/\Delta_{v_{\rm shear}}$ and $\sigma_{\rm med}/\Delta_{\sigma_{\rm med}}$ values are 22 and 73, respectively.
These values are comparable to those of \citet{Herenz2016}'s observations, whose spectral resolutions and SN ratios of H$\alpha$ are similar to those of our observations.
\citet{Herenz2016}'s $v_{\rm shear}/\Delta_{v_{\rm shear}}\sim22$ and $\sigma_{\rm med}/\Delta_{\sigma_{\rm med}}\sim108$, respectively.
}

\subsection{\tcra{Parametric Method}} \label{subsec:galpak}
\begin{figure*}[t]
    \centering
    \includegraphics[width=18.0cm]{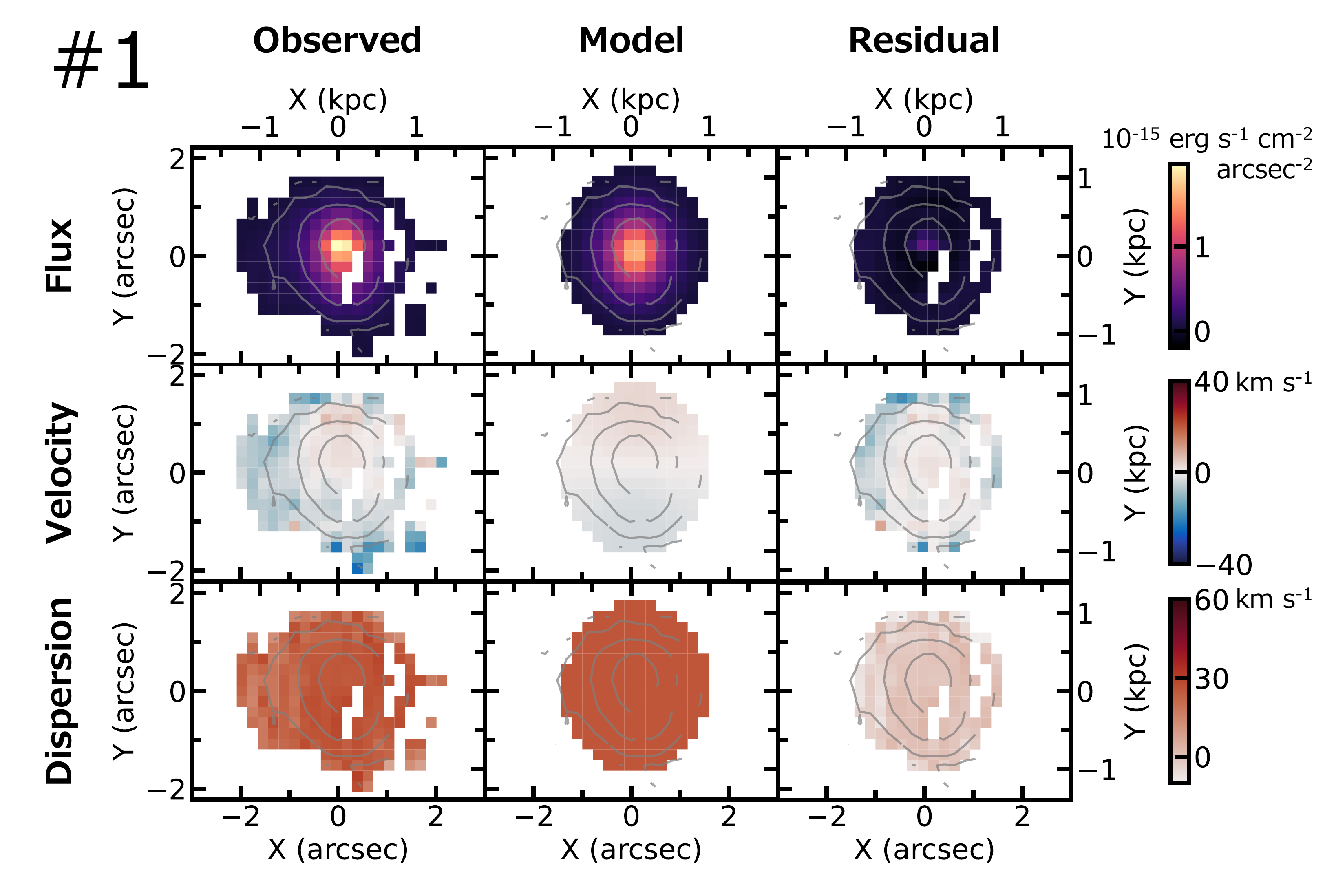}
    \caption{\tcra{GalPaK$^{\rm 3D}$ fitting of J1631+4426. The top, middle, and bottom panels show H$\alpha$ flux, velocity, and velocity-dispersion maps, respectively. The left, center, and right panels illustrate observed, model, and residual ($={\rm observed}-{\rm model}$) maps, respectively. Note that the velocity maps in this figure denote relative velocities with respect to the systemic redshift obtained with GalPaK$^{\rm 3D}$ (Section \ref{subsec:galpak}).}}
    \label{fig:galpak}
\end{figure*}

\begin{table}[t]
    \begin{center}
    \caption{\tcra{Morphological and additional kinematic properties of the 6 EMPGs}}
    \label{tab:add}
    \begin{tabular}{cccccc} \hline \hline
		\# & Name & $r_{\rm e,H\alpha}$ & $i$ & P.A. & $r_{\rm t}$ \\
		& & pc & deg & deg & pc \\
		(1) & (2) & (3) & (4) & (5) & (6) \\ \hline
		1 & J1631+4426 & $248$ & $44$ & $178$ & $61$ \\
		2 & IZw18NW & $149$ & $72$ & $68$ & $339$ \\
		3 & SBS0335$-$052E & $211$ & $25$ & $320$ & $258$ \\
		4 & HS0822+3542 & $33$ & $41$ & $91$ & $1.1$ \\
		5 & J1044+0353 & $90$ & $41$ & $159$ & $106$ \\ 
		6 & J2115$-$1734 & $176$ & $43$ & $222$ & $387$ \\ \hline
    \end{tabular}
    \end{center}
    \tablecomments{(1) Number. (2) Name. (3) H$\alpha$ effective radius. (4) Inclination. (5) Position angle. (6) Turnover radius.}
\end{table}

\tcra{We conduct detailed dynamical modeling with the 3D parametric code GalPaK$^{\rm 3D}$ \citep{Bouche2015}.
Constraining morphological and kinematic properties simultaneously from 3D data cubes, GalPaK$^{\rm 3D}$ provides the deprojected maximum rotational velocity ($v_{\rm rot}$) that is irrespective of the inclination.
GalPaK$^{\rm 3D}$ also calculates $\sigma_{0}$ free from the velocity dispersions driven by the self gravity \citep[e.g.,][]{Genzel2008} or mixture of the line-of-sight velocities.}
Because GalPaK$^{\rm 3D}$ does not support rectangular spaxels, we \tcra{divide} \tcrc{the} spaxels in the Y direction \tcra{based on the linear interpolation} so that the pixel scale in the Y direction is the same as that in the X direction.
We have 10 free parameters of XY coordinates \tcra{of the disk center}, systemic redshift, flux, inclination ($i$), position angle, effective radius \tcra{of H$\alpha$ ($r_{\rm e,H\alpha}$)}, turnover radius ($r_{\rm t}$), $v_{\rm rot}$, and intrinsic velocity dispersion $\sigma_{0}$.
We assume that all 6 EMPGs have thick disks with rotation curves of the arctan profiles:
\begin{equation}
    \label{equ:arctan}
    v(r)=v_{\rm rot}\sin(i)\frac{2}{\pi}\arctan(r/r_{\rm t}),
\end{equation}
where $r$ is a radius from the dynamical center.
\tcra{We note that $i$ is determined by the axis ratio and the disk height, where the disk height of the thick-disk model is fixed to $0.15r_{\rm e,H\alpha}$ \citep{Bouche2015}.}
We also assume that \tcrc{the} surface-brightness (SB) profiles follow S{\'e}rsic profiles with S{\'e}rsic indices $n=1$, which are inferred from $i$-band SB profiles (Paper III).
We have confirmed that these assumptions do not change $v_{\rm rot}$ and $\sigma_{0}$ much.
We also use a point-spread function (PSF) obtained from standard stars with a Moffat profile whose FWHM and power index are $0.\!\!\arcsec7$ and 2.5, respectively.
Figure \ref{fig:galpak} summarizes fitting results.
Table \ref{tab:kin} lists kinematic properties extracted by the GalPaK$^{\rm 3D}$ analysis.
\tcrf{
We estimate the errors of the $v_{\rm rot}$ and $\sigma_{0}$ values by carrying out the Monte Carlo simulations (Section \ref{subsec:meas}).
}
\tcre{We note that $v_{\rm rot}$ can be regarded as an upper limit of the actual rotation velocity because the small-scale velocity differences in the complex velocity field can mimic rotation patterns (see also Section \ref{subsec:shear}).}

\subsection{Mass Profile} \label{subsec:mass}
\begin{figure*}[t]
    \centering
    \includegraphics[width=18.0cm]{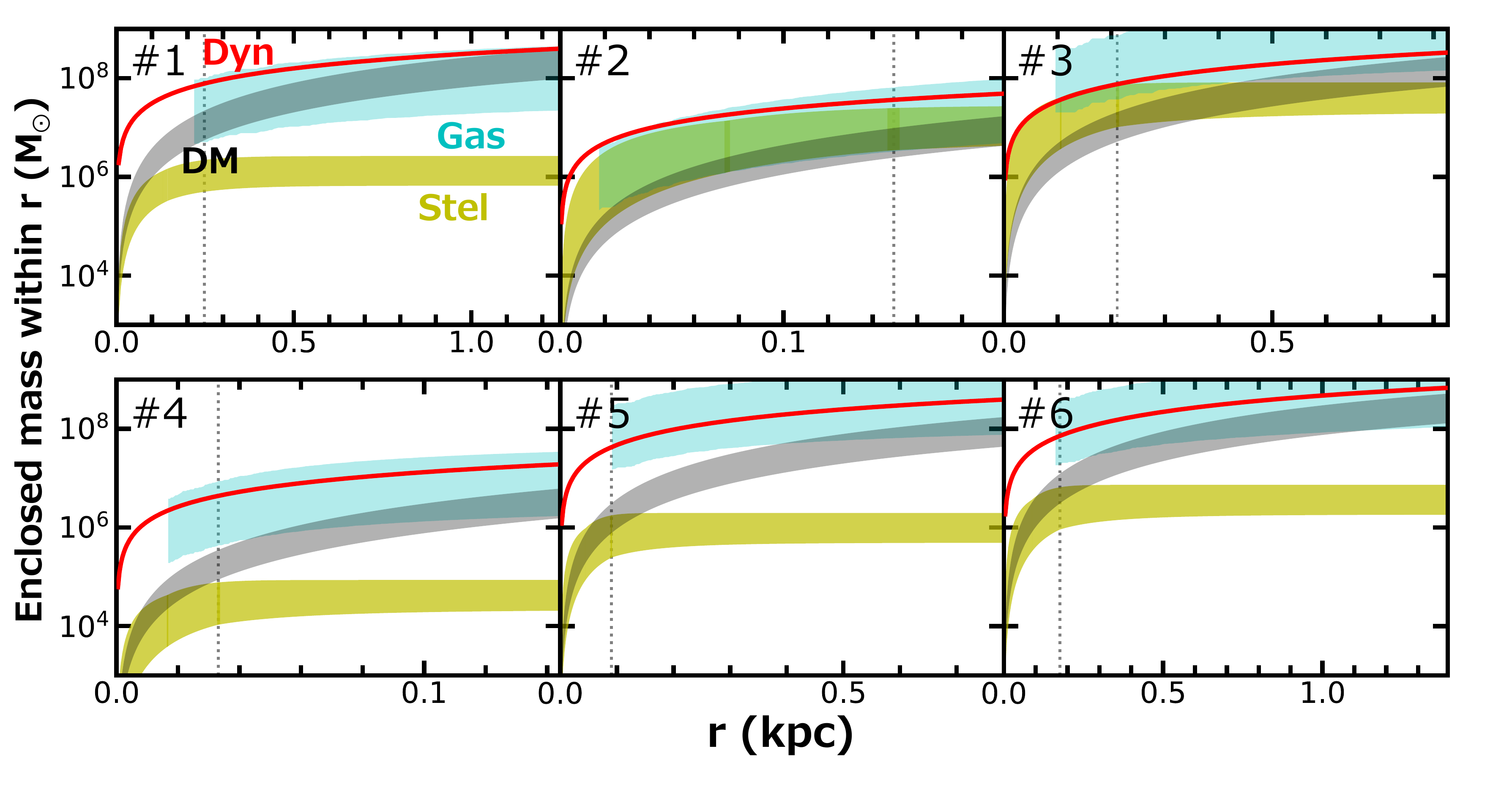}
    \caption{\tcra{Enclosed} mass profiles of the 6 EMPGs. The red, yellow, cyan, and black curves represent dynamical, stellar, gas, and dark-matter mass profiles, respectively. \tcra{The vertical dotted lines show $r_{\rm e,H\alpha}$.} \tcrb{The edge of the plots correspond to the outer most radii used for the kinematic analysis.}}
    \label{fig:profile}
\end{figure*}

\begin{table*}[t]
    \begin{center}
    \caption{\tcrf{Enclosed dynamical, stellar, gas, and DM masses, gas mass fraction, and Toomre $Q$}}
    \label{tab:mass}
    \begin{tabular}{cccccccc} \hline \hline
		\# & Name & $\log[M_{\rm dyn}(<r_{\rm e,H\alpha})]$ & $\log[M_{*}(<r_{\rm e,H\alpha})]$ & $\log[M_{\rm gas}(<r_{\rm e,H\alpha})]$ & $\log[M_{\rm DM}(<r_{\rm e,H\alpha})]$ & $f_{\rm gas}$ & $Q$ \\
		& & $M_{\odot}$ & $M_{\odot}$ & $M_{\odot}$ & $M_{\odot}$ & & \\
		(1) & (2) & (3) & (4) & (5) & (6) & (7) & (8) \\ \hline
		1 & J1631+4426 & $7.90\pm0.06$ & $6.0\pm0.3$ & $7.0^{+1.0}_{-0.3}$ & $7.1\pm0.3$ & $0.91^{+0.06}_{-0.34}$ & $5.1^{+2.1}_{-1.6}$ \\
		2 & IZw18NW & $7.56\pm0.02$ & $7.0\pm0.3$ & $6.8^{+1.0}_{-0.3}$ & $6.7\pm0.3$ & $0.42^{+0.38}_{-0.32}$ & $11.6^{+56.7}_{-3.2}$ \\
		3 & SBS0335$-$052E & $7.88\pm0.02$ & $7.4\pm0.3$ & $7.9^{+1.0}_{-0.3}$ & $7.0\pm0.3$ & $0.74^{+0.17}_{-0.51}$ & $2.6^{+4.7}_{-0.7}$ \\
		4 & HS0822+3542 & $6.64\pm0.03$ & $4.4\pm0.3$ & $5.9^{+1.0}_{-0.3}$ & $5.2\pm0.3$ & $0.97^{+0.03}_{-0.07}$ & $5.4^{+0.2}_{-2.5}$ \\
		5 & J1044+0353 & $7.63\pm0.07$ & $5.8\pm0.3$ & $7.5^{+1.0}_{-0.3}$ & $6.2\pm0.3$ & $0.98^{+0.01}_{-0.09}$ & $3.1^{+3.2}_{-0.0}$ \\ 
		6 & J2115$-$1734 & $7.86\pm0.06$ & $6.4\pm0.3$ & $7.6^{+1.0}_{-0.3}$ & $6.8\pm0.3$ & $0.94^{+0.04}_{-0.18}$ & $1.9^{+0.5}_{-0.6}$ \\ \hline
    \end{tabular}
    \end{center}
    \tablecomments{(1) Number. (2) Name. (3) Dynamical mass \tcra{enclosed in $r_{\rm e,H\alpha}$}. (4) Stellar mass \tcra{enclosed in $r_{\rm e,H\alpha}$}. (5) Gas mass \tcra{enclosed in $r_{\rm e,H\alpha}$}. (6) Dark-matter mass \tcra{enclosed in $r_{\rm e,H\alpha}$}. (7) Gas mass fraction \tcra{within $r_{\rm e,H\alpha}$}. (8) \tcra{Global} Toomre $Q$ parameter (Section \ref{subsec:toomre}).}
\end{table*}
The best-fit disk models obtained in Section \ref{subsec:galpak} provide \tcra{an estimate of the radial profiles for} dynamical masses $M_{\rm dyn}$.
The dynamical mass $M_{\rm dyn}$ is expected to be a sum of stellar mass $M_{*}$, gas mass $M_{\rm gas}$, dark-matter (DM) mass $M_{\rm DM}$, and dust mass $M_{\rm dust}$.
\tcra{Note} that $M_{\rm dust}$ is negligible in all 6 EMPGs because of their low $E(B-V)$ values (e.g., Paper I).
Hereafter, we derive radial profiles of $M_{\rm dyn}$, $M_{*}$, $M_{\rm gas}$, and $M_{\rm DM}$.

\subsubsection{Dynamical mass profile} \label{subsubsec:dyn}
We derive $M_{\rm dyn}$ enclosed within the radius $r$ from the equation
\begin{multline}
    \label{equ:mdyn}
    M_{\rm dyn}(<r)=2.33\times10^{5}\left(\frac{r}{\rm kpc}\right)\times \\
    \left[\left(\frac{v(r)}{\rm km\ s^{-1}}\right)^{2}+2\left(\frac{\sigma_{0}}{\rm km\ s^{-1}}\right)^{2}\right]\ M_{\odot}
\end{multline}
\tcra{under the assumption of the virial equilibrium.}
Figure \ref{fig:profile} presents mass profiles of all 6 EMPGs.
The red curves correspond to $M_{\rm dyn}(<r)$.
Dynamical masses within $r_{\rm e}$ are listed in Table \ref{tab:mass}.

\subsubsection{Stellar mass profile} \label{subsubsec:stel}
The stellar masses of 4 out of the 6 EMPGs (J1631+4426, J2115$-$1734, IZw18NW, and SBS0335$-$052E) are drawn from the literature (Paper I; \citealt{Izotov1998}; \citealt{Izotov2009}). 
\tcra{For the other two targets (J1044+0353 and HS0822+3542), we provide here an estimate of their stellar mass.}

We use the spectral energy distribution (SED) interpretation code of {\sc beagle} \citep{Chevallard2016}.
The {\sc beagle} code calculates both the stellar continuum and the nebular emission using the stellar population synthesis code \citep{Bruzual2003} and the nebular emission library of \citet{Gutkin2016} that are computed with the photoionization code {\sc cloudy} \citep{Ferland2013}.
We fit the SED models to SDSS $ugriz$-band photometries (DR16; \citealt{Ahumada2020}).
We run the {\sc beagle} code with \tcre{4} free parameters of maximum stellar age $t_{\rm max}$, stellar mass $M_{*}$, ionization parameter $U$, and $V$-band optical depth $\tau_{\rm V}$ whose parametric ranges are the same as those adopted in Papers III and IV.
\tcre{We fix metallicities $Z$ to be the gas-phase metallicities $12+\log(\rm O/H)$ listed in Table \ref{tab:fund}.}
We also assume a constant star-formation history and the \citet{Chabrier2003} IMF in the same manner as Papers III and IV.

\tcrb{We note that we conduct the SED fitting with almost the same setting as that of Paper I for J1631+4426 and J2115$-$1734, except for \tcre{$Z$ and} $\tau_{\rm V}$.
\tcre{In Paper I, the $Z$ is treated as a free parameter, and} the $\tau_{\rm V}$ is fixed to be 0, while it \tcrc{has no} impact on $M_{*}$ of J1631+4426 and J2115$-$1734 because both 2 EMPGs are dust-poor (Paper I).}
\tcra{We also confirm that the stellar masses of the 6 EMPGs are different only by $\lesssim0.3$ dex from those derived from $i$-band magnitudes based on the same mass-to-light ratio.
We include this systematic \tcrd{error} of 0.3 dex into the \tcrd{total} error of the $M_{*}$ profile.}

To obtain $M_{*}$ profiles, we assume that \tcra{azimuthally-averaged} $M_{*}$ distributions of the EMPGs follow S{\'e}rsic profiles.
\tcra{J1631+4426 has an $i$-band effective radius ($r_{{\rm e},i}$) of $137^{+9}_{-7}$ pc and an $i$-band S{\'e}rsic index ($n_{i}$) of $1.08^{+0.15}_{-0.13}$ (Paper III).
Because the $i$-band surface-brightness distribution is expected to trace the $M_{*}$ distribution well (Paper III), we assume that J1631+4426 has a stellar effective radius ($r_{\rm e,*}$) and a stellar S{\'e}rsic index ($n_{*}$) equal to $r_{{\rm e},i}$ and $n_{i}$, respectively.
We also assume that the other 5 EMPGs have $r_{\rm e,*}$ within the range from $r_{\rm e,H\alpha}$/2 to $r_{\rm e,H\alpha}$ because $r_{{\rm e},i}$ of J1631+4426 is $\sim2$ times smaller than $r_{\rm e,H\alpha}$ of J1631+4426 (see Table \ref{tab:add})\footnote{\tcra{This result also suggests that EMPGs would have H$\alpha$ halos as discussed in \citet{Herenz2017}.}}.
\tcra{We confirm that the assumed $r_{\rm e,*}$ of the 5 EMPGs are comparable to $r_{{\rm e},i}$ of EMPGs (Paper III).}
The 5 EMPGs are assumed to have $n_{*}$ within the range from 0.7 to 1.7 inferred from the typical $n_{i}$ value of EMPGs (Paper III).
We include these uncertainties of $r_{\rm e,*}$ and $n_{*}$ into the error of the $M_{*}$ profile.}
The yellow shaded regions in Figure \ref{fig:profile} represent cumulative $M_{*}$ profiles with their uncertainties.
Stellar masses within $r_{\rm e}$ are listed in Table \ref{tab:mass}.

\subsubsection{Gas mass profile} \label{subsubsec:gas}
We obtain $M_{\rm gas}$ distributions from the H$\alpha$ flux distributions, using the Kennicutt-Schmidt law (Section \ref{subsec:mask}).
However, \tcra{observational studies \citep[e.g.,][]{Shi2014} have reported that some EMPGs have $\Sigma_{\rm gas}$ values $\sim1$ dex larger than those inferred from the Kennicutt-Schmidt law, which can be interpreted as the lack of metals suppressing efficient gas cooling and succeeding star formation \citep[e.g.,][]{Ostriker2010,Krumholz2013}.}
\tcrb{Given the uncertainty of the Kennicutt-Schmidt law at the low-metallicity end,} we add this 1 dex upper error to the original scatter of the Kennicutt-Schmidt law (\tcrb{$\sim0.3$ dex}; \citealt{Kennicutt1998b}).
The cyan shaded regions in Figure \ref{fig:profile} indicate cumulative $M_{\rm gas}$ profiles with their uncertainties.
\tcra{We note that the {\sc H\,i} observations of \citet{Lelli2012} and \citet{Pustilnik2001} have reported that IZw18NW and SBS0335$-$052E have $M_{\rm gas}\sim1\times10^{8}$ and $\sim1\times10^{9}\ M_{\odot}$ within wide scales of $\sim0.2$ and $\sim3$ kpc, respectively, which are consistent with the extrapolations of the $M_{\rm gas}$ profiles that we derive.}
Gas masses within $r_{\rm e}$ are listed in Table \ref{tab:mass}.
\tcra{We obtain gas mass fractions at $r_{\rm e,H\alpha}$ from the following equation $f_{\rm gas}=M_{\rm gas}(<r_{\rm e,H\alpha})/[M_{\rm gas}(<r_{\rm e,H\alpha})+M_{*}(<r_{\rm e,H\alpha})]$.}

\subsubsection{Dark-matter mass profile} \label{subsubsec:dm}
We estimate $M_{\rm DM}$ profiles under the assumption that the $M_{\rm DM}$ (density) profile follows the NFW profile \citep{Navarro1996}.
At the virial radius $r_{200}$ \tcrd{within which} the spherically-averaged mass density \tcrd{is} 200 times the critical density $\rho_{\rm c}$, the NFW profile can be described by
\begin{equation}
    \label{equ:nfw}
    \rho(r)=\rho_{\rm c}\frac{\delta}{c_{200}x(1+c_{200}x)^{2}},
\end{equation}
where
\begin{equation}
    \label{equ:delta}
    \delta=\frac{200}{3}\frac{c_{200}^{3}}{\ln(1+c_{200})-c_{200}/(1+c_{200})}
\end{equation}
and $x=r/r_{200}$.
\tcra{The \tcrd{concentration} parameter $c_{200}$ \tcrd{is defined as the ratio $r_{200}/r_{\rm d}$ controlling where the slope equals $-2$ as it changes from $-3$ at large radii to a central value of $-1$}.}
We obtain $r_{200}$ from the virial mass $M_{200}$ using the equation
\begin{equation}
    \label{equ:vir}
    r_{200}=\left[\frac{3M_{200}}{4\pi(200\rho_{\rm c})}\right]^{1/3}.
\end{equation}
We derive $M_{200}$ from an empirical stellar-to-halo mass ratio of dwarf galaxies \citep{Brook2014} of
\begin{equation}
    \label{equ:brook}
    M_{200}=7.96\times10^{7}\left(\frac{M_{*}}{M_{\odot}}\right)^{1/3.1}\ M_{\odot}.
\end{equation}
We note that the scatter of the observed stellar-to-halo mass ratios of dwarf galaxies around Equation \ref{equ:brook} is $\sim0.3$ dex \citep{Prole2019a}. 
We include this uncertainty into the error of the $M_{\rm DM}$ profile. 
We also obtain $c_{200}$ from $M_{\rm 200}$ using a halo mass-concentration relation for the Planck cosmology \citep{Dutton2014}:
\begin{equation}
    \label{equ:dutton}
    \log(c_{200})=0.905-0.101\log(M_{200}/10^{12}h^{-1}M_{\odot}),
\end{equation}
where $h=H_{0}/(100\ {\rm km}\ {\rm s}^{-1}\ {\rm Mpc}^{-1})$\footnote{\tcrb{We use $h=0.7$ for consistency with our analysis based on the standard $\Lambda{\rm CDM}$ cosmology (Section \ref{sec:intro}), while we confirm that $h=0.67$ based on the Planck cosmology does not change our conclusion.}}.
The gray shaded regions in Figure \ref{fig:profile} denote $M_{\rm DM}$ profiles with their uncertainties.
DM masses within $r_{\rm e}$ are listed in Table \ref{tab:mass}.

\section{Results} \label{sec:result}
\subsection{\tcra{Rotation and Dispersion}} \label{subsec:vsig}
\begin{figure*}[t]
    \centering
    \includegraphics[width=18.0cm]{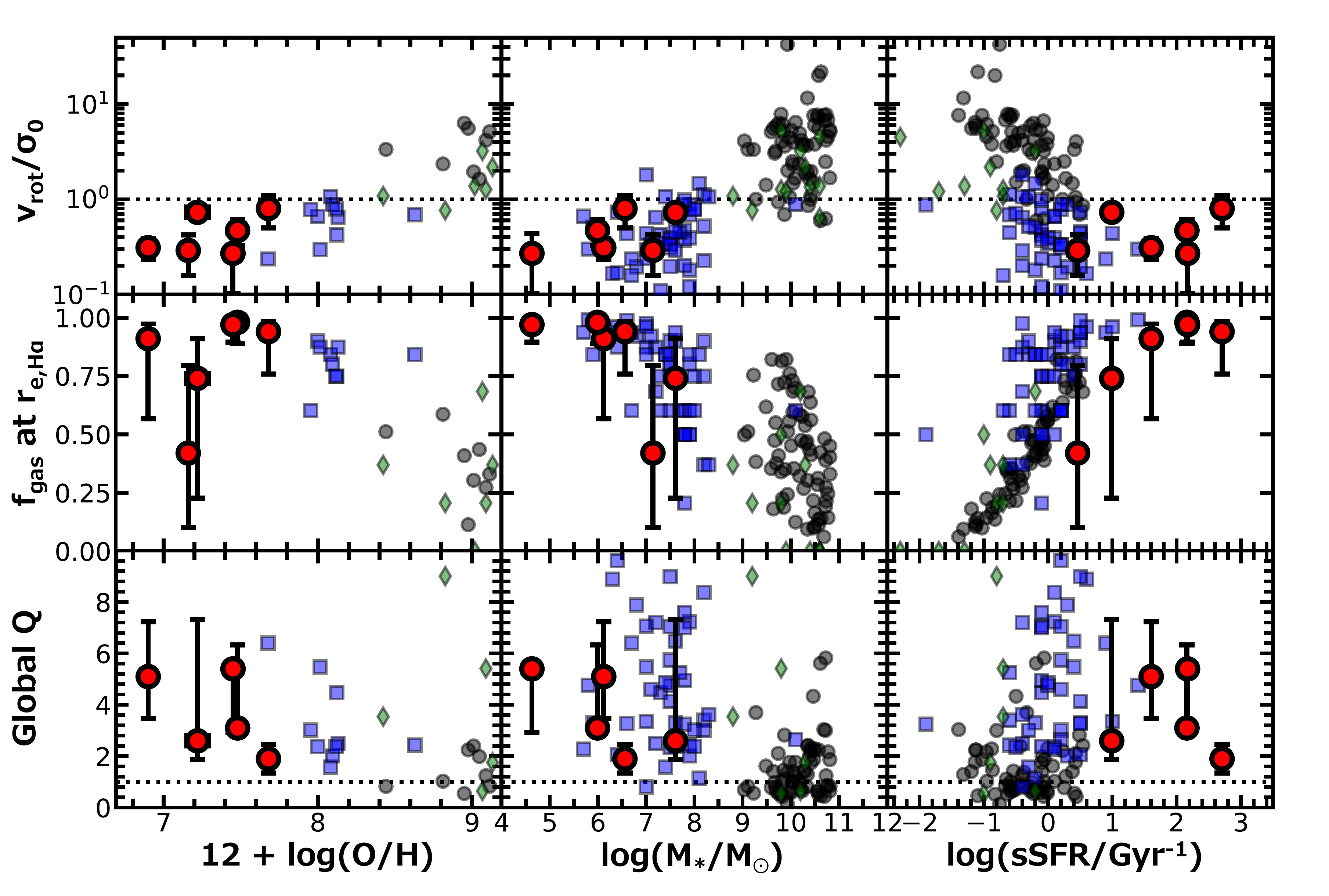}
    \caption{Diagrams of $v_{\rm rot}/\sigma_{0}$, $f_{\rm gas}$, and global $Q$ as functions of $12+\log(\rm O/H)$, $M_{*}$, and sSFR. \tcrb{The red circles represent the EMPGs, while the blue squares, the green diamonds, and the black circles indicate the SH$\alpha$DE \citep{Barat2020}, SAMI \citep{Barat2019}, and DYNAMO galaxies \citep{Green2014}, respectively.} \tcrf{Because IZw18NW has the large uncertainty of the global $Q$ (see Table \ref{tab:mass}), we exclude the data point of this EMPG from the bottom panels.}}
    \label{fig:main}
\end{figure*}

Table \ref{tab:kin} summarizes \tcra{$v_{\rm shear}$, $\sigma_{\rm med}$}, $v_{\rm rot}$, and $\sigma_{0}$ of the 6 EMPGs.
\tcra{Note that $v_{\rm shear}$ and $\sigma_{\rm med}$ ($v_{\rm rot}$ and $\sigma_{0}$) are based on the non-parametric (parametric) method explained in Section \ref{subsec:shear} (\ref{subsec:galpak}).
We find that all 6 EMPGs have low $v_{\rm shear}$ (5.5--14.3 km s$^{-1}$), high $\sigma_{\rm med}$ (16.9--30.8 km s$^{-1}$), and thus low $v_{\rm shear}/\sigma_{\rm med}$ (0.18--0.58).}
Regarding SBS0335$-$052E, \citet{Herenz2017} also obtain $v_{\rm shear}/\sigma_{\rm med}=0.68$, which is \tcrf{close to our measurement of $v_{\rm shear}/\sigma_{\rm med}=0.58\pm0.02$.}
The 6 EMPGs also have low $v_{\rm rot}$ (4.5--23.4 km s$^{-1}$), high $\sigma_{0}$ (16.6--31.4 km s$^{-1}$), and low $v_{\rm rot}/\sigma_{0}$ (0.27--0.80).
\tcra{Using the 2 different methods, we confirm that all 6 EMPGs are dominated by dispersion (i.e., $v_{\rm shear}/\sigma_{\rm med}$, $v_{\rm rot}/\sigma_{0}<1$).}

The top left panel of Figure \ref{fig:main} shows $v_{\rm rot}/\sigma_{0}$ values of the 6 EMPGs (red circle) as a function of metallicity.
We compare our results with other surveys of H$\alpha$ kinematics of star-forming dwarf galaxies (SH$\alpha$DE; \citealt{Barat2020}) and star-forming galaxies (SAMI; \citealt{Barat2019}; DYNAMO; \citealt{Green2014}), whose metallicities are drawn from the SDSS MPA-JHU catalog \citep{Tremonti2004,Brinchmann2004}.
The SH$\alpha$DE, SAMI, and DYNAMO galaxies have $12+\log(\rm O/H)\sim8$--9.
We find that $v_{\rm rot}/\sigma_{0}$ decreases with decreasing $12+\log(\rm O/H)$.
The top middle and top right panels of Figure \ref{fig:main} show $v_{\rm rot}/\sigma_{0}$ as a function of $M_{*}$ \tcra{and} sSFR, respectively.
\tcre{The SFRs are derived from H$\alpha$ luminosity.
The $M_{*}$ and sSFR potentially have uncertainties of $\sim0.3$ dex under different assumptions such as initial mass functions (cf. Section \ref{subsubsec:stel}).}
Although it should be noted that the EMPGs are biased toward lower metallicities, we also find that $v_{\rm rot}/\sigma_{0}$ decreases as $M_{*}$ decreases and sSFR increases.
These results \tcrc{suggest} that galaxies \tcrd{in earlier stages of the} formation phase \tcrc{may} have lower $v_{\rm rot}/\sigma_{0}$.

\subsection{Mass Fraction} \label{subsec:fgas}
Figure \ref{fig:profile} summarizes radial profiles of $M_{\rm dyn}$ (red), $M_{*}$ (yellow), $M_{\rm gas}$ (cyan), and $M_{\rm DM}$ (black).
We find that the 4 EMPGs other than IZw18NW or SBS0335$-$052E have the $M_{*}$ profiles \tcra{$\sim2$ dex below} the $M_{\rm dyn}$ profiles within radii up to several times $r_{\rm e,H\alpha}$, which means that the 4 EMPGs are dominated by $M_{\rm gas}$ or $M_{\rm DM}$ on galactic scales.
\tcra{On the other hand, IZw18NW and SBS0335$-$052E have the $M_{*}$ profiles comparable to the $M_{\rm dyn}$ profiles.}
We also find that the $M_{\rm dyn}$ profiles of all 6 EMPGs can be explained by the $M_{\rm gas}$ profiles within the uncertainties (see Section \ref{subsubsec:gas}).
We confirm that the $M_{\rm DM}$ profiles of all 6 EMPGs are \tcra{$\sim1$ dex below} the $M_{\rm dyn}$ profiles within radii up to several times $r_{\rm e,H\alpha}$.
We thus conclude that the masses of the 4 EMPGs except for IZw18NW and SBS0335$-$052E are dominated by $M_{\rm gas}$ on galactic scales.
\tcra{We note that IZw18NW and SBS0335$-$052E indeed have large $M_{\rm gas}$ values of $\sim1\times10^{8}$ and $\sim1\times10^{9}\ M_{\odot}$ \tcrc{inferred from the H\,{\sc i} observations} within $\sim0.2$ and $\sim3$ kpc, respectively (Section \ref{subsubsec:gas}).
Within these larger scales, we can say that both IZw18NW and SBS0335$-$052E are gas-rich (i.e., $f_{\rm gas}\sim1$).}
\tcre{This conclusion is consistent with those in \citet{Pustilnik2020a,Pustilnik2020b,Pustilnik2021} based on H\,{\sc i} observations.}
It should also be noted that Equations \ref{equ:vir} and \ref{equ:brook} suggest that EMPGs with $\sim10^{6}\ M_{\odot}$ have $M_{200}\sim7\times10^{9}\ M_{\odot}$ at $r_{200}\sim30$ kpc, whereas we can only observe the area within at most $\sim1$ kpc. 
\tcra{Therefore, it is natural that $M_{\rm DM}$ has negligible effects on the mass profile in this study.}

The middle left panel of Figure \ref{fig:main} shows $f_{\rm gas}$ of the 6 EMPGs as a function of metallicity.
\tcra{Except for IZw18NW and SBS0335$-$052E with large $f_{\rm gas}$ uncertainties,} we find that the EMPGs \tcrd{are gas-rich with} high $f_{\rm gas}$ values of $\sim0.9$--1.0.
Comparing with the literature, we find that galaxies with lower metallicities tend to have higher $f_{\rm gas}$ values.
The $f_{\rm gas}$ values also increase \tcrd{at} lower $M_{*}$ and higher sSFR.
These results indicate that galaxies \tcrd{in} the earlier formation phase would have higher $f_{\rm gas}$, which are consistent with \tcra{previous observations \citep[e.g.,][]{Maseda2014} as well as} models of galaxy formation based on the $\Lambda$CDM model \citep[e.g.,][]{Geach2011}.
It should be noted that $M_{\rm gas}$ of the SH$\alpha$DE and DYNAMO galaxies are estimated from Kennicutt-Schmidt law (i.e., correlation between $\Sigma_{\rm SFR}$ and $\Sigma_{\rm gas}$) in the similar way as our analysis (cf. Section \ref{subsubsec:gas}).
Thus, it is natural that we find \tcrc{a} tight correlation between sSFR and $f_{\rm gas}$.

\section{Discussion} \label{sec:dis}
\subsection{Origin of Low $v_{\rm rot}/\sigma_{0}$} \label{subsec:orig}
In Section \ref{subsec:vsig}, we \tcra{report} that all 6 EMPGs have low $v_{\rm rot}/\sigma_{0}<1$.
We also find that $v_{\rm rot}/\sigma_{0}$ decreases with decreasing $12+\log(\rm O/H)$, decreasing $M_{*}$, and increasing sSFR (Figure \ref{fig:main}).
Below, we investigate \tcra{well-discussed \tcrc{three}} contributors \tcra{\citep[e.g.,][]{Glazebrook2013,Barat2020} to \tcrc{such a}} low $v_{\rm rot}/\sigma_{0}<1$.

\subsubsection{Thermal expansion} \label{subsubsec:therm}
The first possible contributor \tcra{to the low $v_{\rm rot}/\sigma_{0}$} is the thermal expansion of H\,{\sc ii} regions \citep[e.g.,][]{Krumholz2018,Fukushima2021,Fukushima2022}. 
We estimate a velocity dispersion of the thermal expansion ($\sigma_{\rm th}$) \tcra{from the \tcre{line-of-sight component} of the Maxwellian velocity distribution (\tcre{$\sigma_{\rm th}=\sqrt{kT_{\rm e}/m}$}; e.g., \citealt{Chavez2014,Pillepich2019})}, where $k$, $T_{\rm e}$, and $m$ represent the Boltzmann constant, the electron temperature, and the hydrogen mass, respectively.
We obtain $\sigma_{\rm th}=\tcre{9.1}$ km s$^{-1}$ under the assumption of $T_{\rm e}=10000$ K, which is consistent with the typical electron temperature of O\,{\sc ii} (i.e., H\,{\sc ii}) regions of the EMPGs (e.g., Paper I). 
Subtracting $\sigma_{\rm th}$ from $\sigma_{0}$ quadratically, we obtain velocity dispersions being free from the thermal line broadening \tcra{($\sigma_{\rm no\_therm}=\tcre{14}$\tcre{--30} km s$^{-1}$)}. 
We confirm that $v_{\rm rot}/\sigma_{\rm no\_therm}$ values are still lower than unity \tcre{(0.31--0.84)} for all 6 EMPGs.
We thus conclude that the thermal expansion cannot explain $v_{\rm rot}/\sigma_{0}<1$.

\subsubsection{Merger/inflow} \label{subsubsec:merg}
The second possible contributor is merger/inflow events \citep[e.g.,][]{Glazebrook2013}.
The merger (inflow) can raise velocity dispersions by tidal heating (releasing potential energies of infalling gas).
This scenario can explain all the trends seen in Figure \ref{fig:main} because we can expect that both \tcra{gas-rich minor} merger and inflow would supply metal-poor gas and trigger succeeding starbursts.
\tcra{Especially for J1631+4426, IZw18NW, SBS0335$-$052E, and J2115$-$1734, we find the velocity differences from the EMPG tails (IZw18SE for IZw18NW) suggestive of merger (Section \ref{sec:map}).}

\subsubsection{\tcra{Stellar} feedback} \label{subsubsec:sf}
The third possible contributor is the stellar feedback \citep[e.g.,][]{Lehnert2009}.
This includes the supernova (SN) feedback \citep{Dib2006} as well as stellar winds and the radiative pressure from young massive stars \citep{MacLow2004}.
\tcrd{Outflowing} gas from SNe and/or young massive stars would raise velocity dispersions.
This stellar feedback scenario can directly explain the decreasing trend between $v_{\rm rot}/\sigma_{0}$ and sSFR (the top right panel of Figure \ref{fig:main}).
Given the expectation that young galaxies (i.e., with high sSFRs) would have low $12+\log(\rm O/H)$ and $M_{*}$, \tcrd{this} scenario can indirectly reproduce the trend that $v_{\rm rot}/\sigma_{0}$ decreases with decreasing $12+\log(\rm O/H)$ and $M_{*}$.
IZw18NW, SBS0335$-$052E, and J1044+0353 \tcrd{especially} show outflow signatures in flux, velocity, and velocity-dispersion maps (see Section \ref{sec:map}), which imply the dominance of the stellar feedback.
We thus conclude that the stellar feedback can also be one of the main contributors of $v_{\rm rot}/\sigma_{0}<1$ at the low-metallicity, low-$M_{*}$, and high-sSFR ends.
Cosmological zoom-in simulations will provide a hint of \tcra{what kind of feedback is the main contributor}.

\subsection{\tcra{Toomre Q Parameter}} \label{subsec:toomre}
\tcra{To compare with other kinematic studies,} we derive the Toomre $Q$ parameter \citep{Toomre1964} of the EMPGs.
\tcre{In general, if the $Q$ value of a rotating disk is greater than unity (i.e., $Q>1$), the disk is thought to be gravitationally stable.
On the other hand, the disk is gravitationally unstable if $Q<1$.}
However, note that it is unclear if this criterion is applicable for the EMPGs because they may not have rotating disks (Section \ref{sec:map}).
\tcra{An average of $Q$ within a disk so called the global $Q$ \citep[e.g.,][]{Aumer2010} is calculated} from the equation
\begin{equation}
    \label{equ:toomre}
    Q=\frac{\sigma_{0}}{v_{\rm rot}}\frac{a}{f_{\rm gas}},
\end{equation}
where $a$ is a parameter with values ranging from 1 to 2 depending on the gas distribution \citep{Genzel2011}.
Here, we adopt $a=\sqrt{2}$ assuming the constant rotational velocity.

\tcrf{Table \ref{tab:mass} lists the global $Q$ of the 6 EMPGs.
We find that all the 6 EMPGs show $Q>1$, albeit one of the 6 EMPGs (IZw18NW) with the large $Q$ uncertainty.
The bottom panels of Figure \ref{fig:main} illustrate the global $Q$ of the EMPGs as a function of metallicity (bottom left), $M_{*}$ (bottom center), and sSFR (bottom right), while we exclude IZw18NW due to its large $Q$ uncertainty.}
We also find that the global $Q$ increases with decreasing $12+\log(\rm O/H)$, decreasing $M_{*}$, and increasing sSFR.
\tcre{However, the large global $Q$ values are inconsistent with the large sSFR values because star-formation activities are not likely to become aggressive in a gravitationally-stable disk.}

\tcre{This inconsistency probably originates from the fact that the global $Q$ parameter is not a reliable indicator for gas-rich galaxies \citep{Romeo2010,Romeo2014}.
Instead, it can be important to investigate if the EMPGs lie on a tight relation based on observables such as H\,{\sc i} angular momentum \citep{Romeo2020a,Romeo2020b}.
We need high-resolution H\,{\sc i} observations to discuss gravitational instability of EMPGs precisely.}


\subsection{Connection to High-$z$ Primordial Galaxies} \label{subsec:highz}
\begin{figure*}[t]
    \centering
    \includegraphics[width=18.0cm]{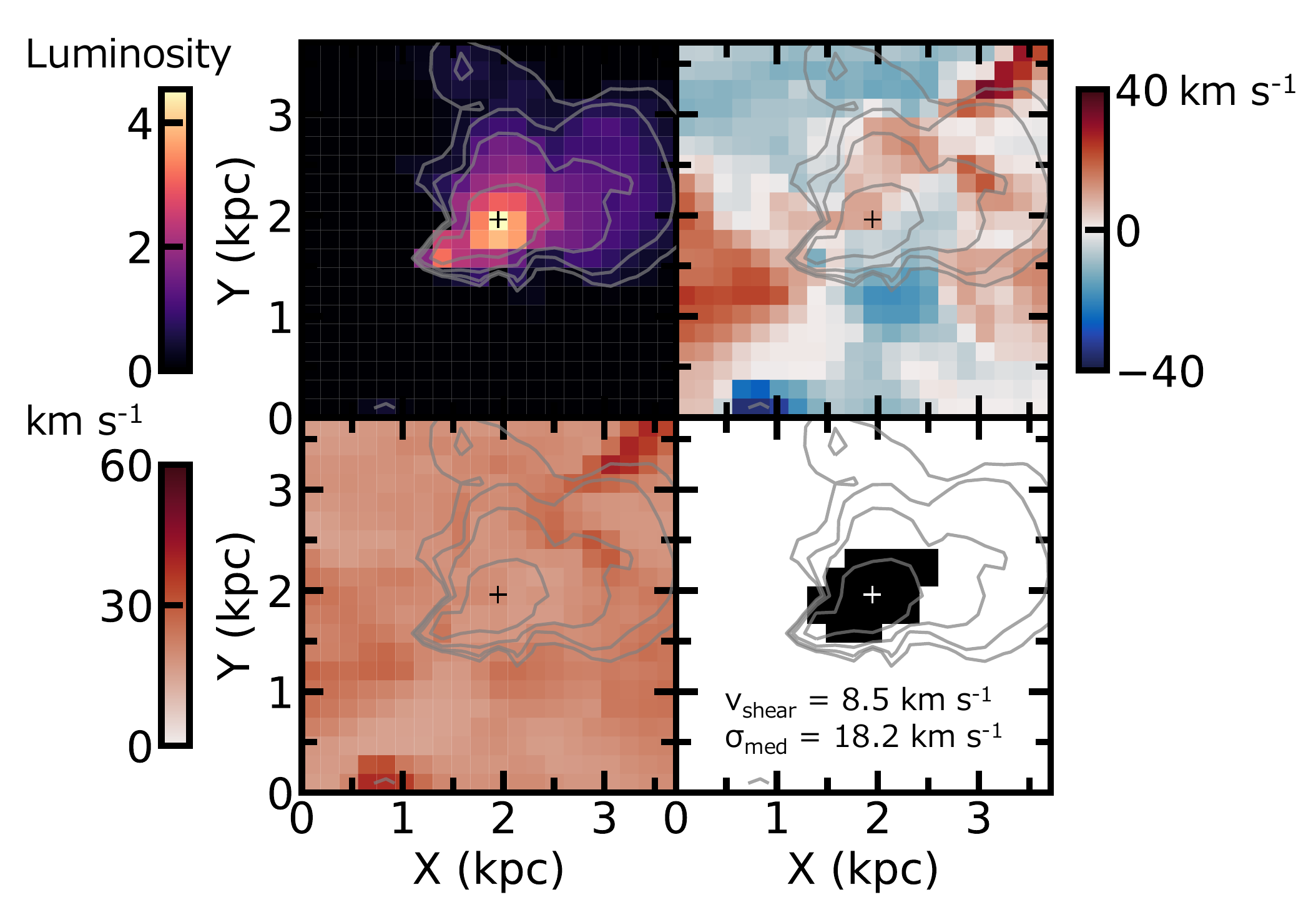}
    \caption{\tcrb{H$\alpha$ luminosity (top left), velocity (top right), velocity-dispersion (bottom left), and mask maps (bottom right) of the $z=7.3$ primordial galaxy in \citet{Wise2014}. \tcrc{The luminosity is in units of $10^{35}$ erg s$^{-1}$ \AA$^{-1}$ in log scale. The maximum luminosity value ($7\times10^{39}$ erg s$^{-1}$ \AA$^{-1}$) corresponds to a flux of $3\times10^{-15}$ erg s$^{-1}$ cm$^{-2}$ \AA$^{-1}$ at $z=0.03125$.} The gray contours illustrate H$\alpha$ luminosity values in the order of 1/16, 1/8, 1/4, and 1/2 of the maximum luminosity value.}}
    \label{fig:wise}
\end{figure*}

The trend that $v_{\rm rot}/\sigma_{0}$ ($f_{\rm gas}$) decreases (\tcrd{increases}) with decreasing (increasing) $12+\log(\rm O/H)$ and $M_{*}$ (sSFR) \tcra{suggests} that primordial galaxies at high redshifts would be \tcra{dispersion-dominated} gas-rich systems.
This \tcra{suggestion} agrees with the decreasing trend of $v_{\rm rot}/\sigma_{0}$ with the redshift reported by \tcrb{both observational \citep[e.g.,][]{Wisnioski2015} and simulation \citep{Pillepich2019} studies}.

\tcrb{Here, we investigate kinematics of simulated primordial galaxies at high redshifts.
In this study, we choose a $z=7.3$ primordial galaxy of \citet{Wise2014}'s cosmological radiation hydrodynamics simulation because the simulated galaxy has a low gas-phase metallicity ($\sim4$\% $Z_{\odot}$), a low stellar mass ($3.8\times10^{6}\ M_{\odot}$), \tcrc{a large $f_{\rm gas}$ ($\sim1$), a large sSFR ($\sim5$ Gyr$^{-1}$),} and a small half-mass radius ($\sim200$ pc), all of which are comparable\footnote{Under the assumption that the half-mass radius is comparable to the rest-frame $i$-band effective radius.} to those of the EMPGs (Sections \ref{sec:sample} and \ref{subsubsec:stel}). 
Wise et al. in prep. (hereafter \tcrg{W23}) extract H$\alpha$ flux, velocity, and velocity-dispersion maps \tcrd{from} the simulated galaxy.
The FoV of the extracted region is 3.71 kpc with the spatial resolution of $3.71/252=0.015$ kpc pixel$^{-1}$.
The spectral resolution of the datacube is 0.1\ \AA.
\tcrg{W23} use {\sc cloudy} \citep{Ferland2013} to derive H$\alpha$ fluxes from the following 5 quantities: Hydrogen number density, temperature, metallicity, incident radiation intensity, and H\,{\sc i} column density, along with enough parametric ranges and number of datapoints (1.8 million points in total).
For each cell in the data cube of the 5 quantities, \tcrg{W23} linearly interpolate the emissivity table in 5D.
To calculate the velocity map, \tcrg{W23} sum the line-of-sight velocities weighted by the H-alpha fluxes.
To calculate the velocity dispersion map, \tcrg{W23} use the same method as \citeauthor{Pillepich2019} (\citeyear{Pillepich2019}; Equation A3), using the H-alpha flux as the weights.
We note that the velocity dispersions include the thermal broadening.
}

\tcrb{We coarsen the data cube to a spatial resolution of 180 pc, similar to the resolution of our observations, to make a \tcrc{more relevant} comparison.
The top left, top right, and bottom left panels of Figure \ref{fig:wise} are H$\alpha$ flux, velocity, and velocity-dispersion maps of the simulated galaxy, respectively.
We find that the simulated galaxy has an irregular morphology with multiple kinematic sub-structures and localized turbulent regions.
These features can be seen in the EMPGs as well (Figure \ref{fig:map}).
Masking out the kinematic sub-structures and the turbulent regions, we also derive kinematics properties of $v_{\rm shear}$ and $\sigma_{\rm med}$ in the same manner as we do for the EMPGs (Sections \ref{subsec:mask} and \ref{subsec:shear}).
We find that the simulated galaxy has a low $v_{\rm shear}$ (8.5 km s$^{-1}$) and a high $\sigma_{\rm med}$ (18.2 km s$^{-1}$), which are comparable to those of the EMPGs ($v_{\rm shear}=5.5$--14.3 km s$^{-1}$, $\sigma_{\rm med}=16.9$--30.8 km s$^{-1}$; see Section \ref{subsec:vsig}).
Consequently, the simulated galaxy has a low $v_{\rm shear}/\sigma_{\rm med}=0.47$ below unity suggesting that the simulated galaxy is dominated by dispersion as well as the EMPGs.
Given that the local EMPGs, analogs of high-$z$ primordial galaxies, and the simulated high-$z$ primordial galaxies are both dispersion-dominated systems, we expect that high-$z$ primordial galaxies are likely to be dispersion-dominated galaxies.}

The forthcoming James Webb Space Telescope (JWST) can \tcra{directly investigate H$\alpha$ kinematics of high-$z$ primordial galaxies}.
Paper IV has simulated H$\alpha$ fluxes of primordial galaxies with $M_{*}=10^{6}\ M_{\odot}$ at redshifts ranging from 0 to 10.
Comparing the H$\alpha$ fluxes with the limiting flux of JWST/NIRSpec, we estimate that NIRSpec can detect H$\alpha$ fluxes of primordial galaxies with $M_{*}=10^{6}\ M_{\odot}$ at $z\lesssim1$ without gravitational lensing.
With \tcrd{a} $\sim2$ dex magnification \tcrd{from} gravitational lensing, NIRSpec can detect H$\alpha$ fluxes of primordial galaxies with $M_{*}=10^{6}\ M_{\odot}$ at $z\sim\tcrb{7}$.
We infer from this estimation that NIRSpec could observe primordial galaxies with $M_{*}\sim10^{7}\ M_{\odot}$ at $z\sim7$ with a realistic magnification of $\sim1$ dex.
The top middle panel of Figure \ref{fig:main} shows that most of galaxies with $M_{*}\sim10^{7}\ M_{\odot}$ already have low $v_{\rm rot}/\sigma_{0}<1$ and high $f_{\rm gas}>0.5$.
Lensing cluster surveys using JWST such as GLASS (PI: T. Treu) and CANUCS (PI: C. Willott) will potentially pinpoint low-mass galaxies with $M_{*}\sim10^{7}\ M_{\odot}$ at $z\sim7$, and follow-up IFU observations with JWST may identify \tcra{dispersion-dominated} gas-rich galaxies.

\section{Summary} \label{sec:sum}
We present kinematics of 6 local extremely metal-poor galaxies (EMPGs) with low metallicities ($0.016-0.098\ Z_{\odot}$) and low stellar masses ($10^{4.7}-10^{7.6} M_{\odot}$; Section \ref{sec:sample}).
Taking deep medium-high resolution ($R\sim7500$) integral-field spectra with 8.2-m Subaru (Section \ref{sec:obs}), we resolve the small inner velocity gradients and dispersions of the EMPGs with H$\alpha$ emission. 
Carefully masking out sub-structures originated by inflow and/or outflow,
we fit 3-dimensional disk models to the observed H$\alpha$ flux, velocity, and velocity-dispersion maps (Sections \ref{sec:map} and \ref{sec:analysis}).
All the EMPGs show rotational velocities ($v_{\rm rot}$) of 5--23 km s$^{-1}$ smaller than the velocity dispersions
($\sigma_{0}$) of 17--31 km s$^{-1}$, indicating dispersion-dominated systems with small ratios of $v_{\rm rot}/\sigma_{0}=0.29-0.80$ (Section \ref{subsec:vsig}) that can be explained by turbulence driven by inflow and/or outflow (Section \ref{subsec:orig}).
Except for two EMPGs with large uncertainties, we find that the EMPGs have very large gas-mass fractions of $f_{\rm gas}\simeq 0.9-1.0$ (Section \ref{subsec:fgas}). Comparing our results with other H$\alpha$ kinematics studies, we find that $v_{\rm rot}/\sigma_{0}$ ($f_{\rm gas}$) decreases (increases) with decreasing metallicity.
\tcrb{We compare numerical simulations of first-galaxy formation and identify that the simulated high-$z$ ($z\sim 7$) forming galaxies have gas-fractions and dynamics similar to the observed EMPGs. Our EMPG observations and the simulations suggest that primordial galaxies are gas-rich dispersion-dominated systems\tcrc{, which} would be identified by the forthcoming JWST observations at $z\sim 7$ (Section \ref{subsec:highz}).}
\\

We are grateful to Yoshiaki Sofue, \tcre{Alessandro Romeo, and Roberto Terlevich} for having useful discussions.
We also thank the staff of Subaru Telescope for their help with the observations. 
This research is based on data collected at the Subaru Telescope, which is operated by the National Astronomical Observatory of Japan (NAOJ).
We are honored and grateful for the opportunity of observing the Universe from Maunakea, which has the cultural, historical, and natural significance in Hawaii.
The Hyper Suprime-Cam (HSC) collaboration includes the astronomical communities of Japan and Taiwan, and Princeton University.
The HSC instrumentation and software were developed by the NAOJ, the Kavli Institute for the Physics and Mathematics of the Universe (Kavli IPMU), the University of Tokyo, the High Energy Accelerator Research Organization (KEK), the Academia Sinica Institute for Astronomy and Astrophysics in Taiwan (ASIAA), and Princeton University.
Based on data collected at the Subaru Telescope and retrieved from the HSC data archive system, which is operated by Subaru Telescope and Astronomy Data Center at NAOJ.
This work was supported by the joint research program of the Institute for Cosmic Ray Research (ICRR), University of Tokyo. 
\tcrb{
Y.I., K. Nakajima, Y.H., T.K., and M. Onodera are supported by JSPS KAKENHI Grant Nos. 21J20785, 20K22373, 19J01222, 18J12840, and 21K03622, respectively.
K.H. is supported by JSPS KAKENHI Grant Nos. 20H01895, 21K13909, and 21H05447.
Y.H. is supported by JSPS KAKENHI Grant Nos. 21J00153, 20K14532, 21H04499, 21K03614, and 22H01259.
H.Y. is supported by MEXT / JSPS KAKENHI Grant Number 21H04489 and JST FOREST Program, Grant Number JP-MJFR202Z.
J.H.K acknowledges the support from the National Research Foundation of Korea (NRF) grant, \tcrf{No. 2020R1A2C3011091 and} No. 2021M3F7A1084525, funded by the Korea government (MSIT).
This work has been supported by the Japan Society for the Promotion of Science (JSPS) Grants-in-Aid for Scientific Research (19H05076 and 21H01128).
This work has also been supported in part by the Sumitomo Foundation Fiscal 2018 Grant for Basic Science Research Projects (180923), and the Collaboration Funding of the Institute of Statistical Mathematics ``New Development of the Studies on Galaxy Evolution with a Method of Data Science''.
The Cosmic Dawn Center is funded by the Danish National Research Foundation under grant No. 140. 
S.F. acknowledges support from the European Research Council (ERC) Consolidator Grant funding scheme (project ConTExt, grant No. 648179). 
This project has received funding from the European Union's Horizon 2020 research and innovation program under the Marie Sklodowska-Curie grant agreement No. 847523 ``INTERACTIONS''.
This work is supported by World Premier International Research Center Initiative (WPI Initiative), MEXT, Japan, as well as KAKENHI Grant-in-Aid for Scientific Research (A) (15H02064, 17H01110, 17H01114, 20H00180, and 21H04467) through Japan Society for the Promotion of Science (JSPS). 
This work has been supported in part by JSPS KAKENHI Grant Nos. JP17K05382, JP20K04024, and JP21H04499 (K. Nakajima).}
\tcre{This research was supported by a grant from the Hayakawa Satio Fund awarded by the Astronomical Society of Japan.}
JHW acknowledges support from NASA grants NNX17AG23G, 80NSSC20K0520, and 80NSSC21K1053 and NSF grants OAC-1835213 and AST-2108020.
\software{\tcre{FOCAS IFU pipeline \citep{Ozaki2020}, PyRAF \citep{STScI2012}, astropy \citep{Astropy2013,Astropy2018}, GalPaK$^{\rm 3D}$ \citep{Bouche2015}, beagle \citep{Chevallard2016}, cloudy \citep{Ferland2013}, yt \citep{Turk2011}}}

\appendix
\begin{figure*}[t]
    \centering
    \includegraphics[width=18.0cm]{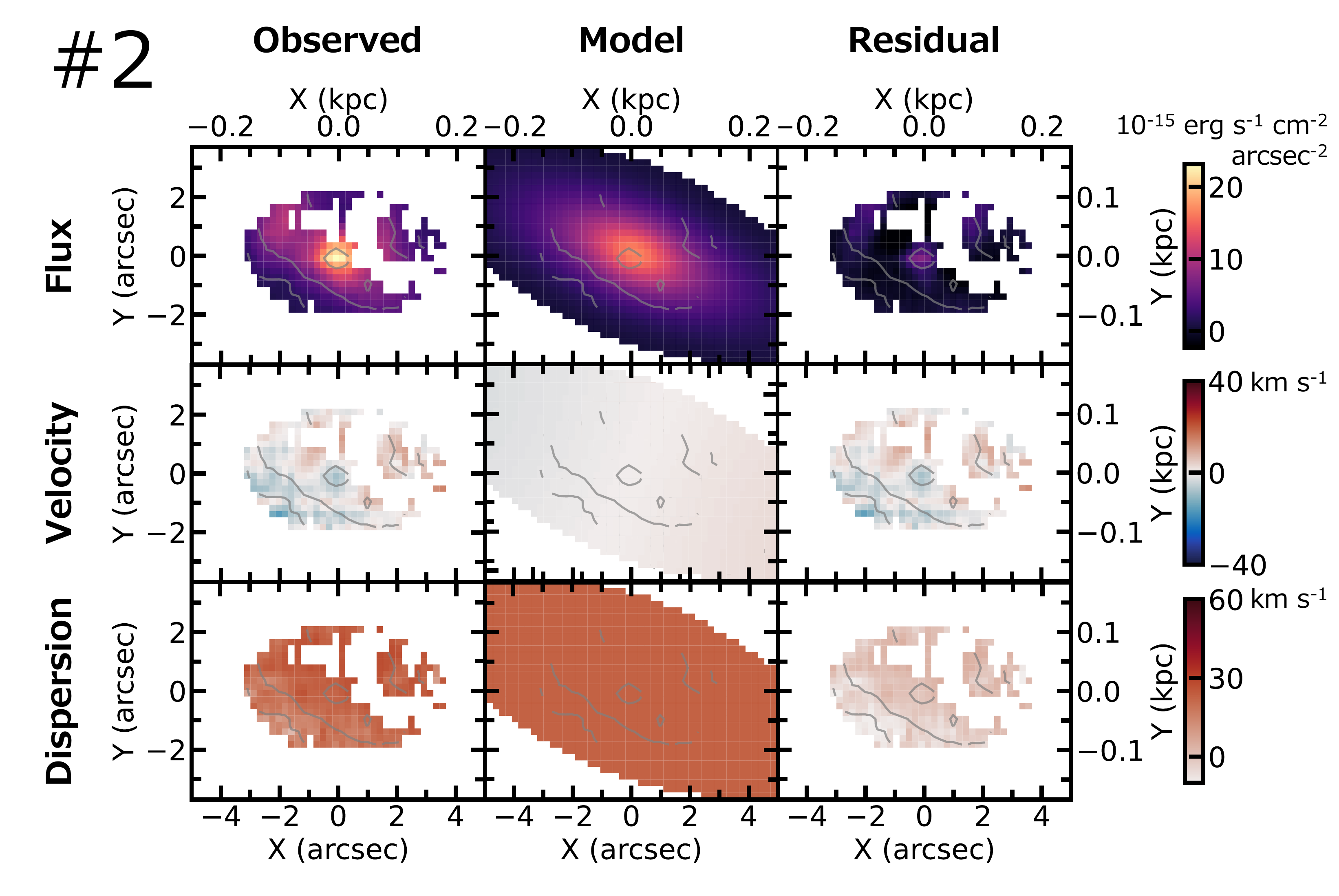}
    \caption{GalPaK$^{\rm 3D}$ result of IZw18NW.}
    \label{fig:galpak_2}
\end{figure*}

\begin{figure*}[t]
    \centering
    \includegraphics[width=18.0cm]{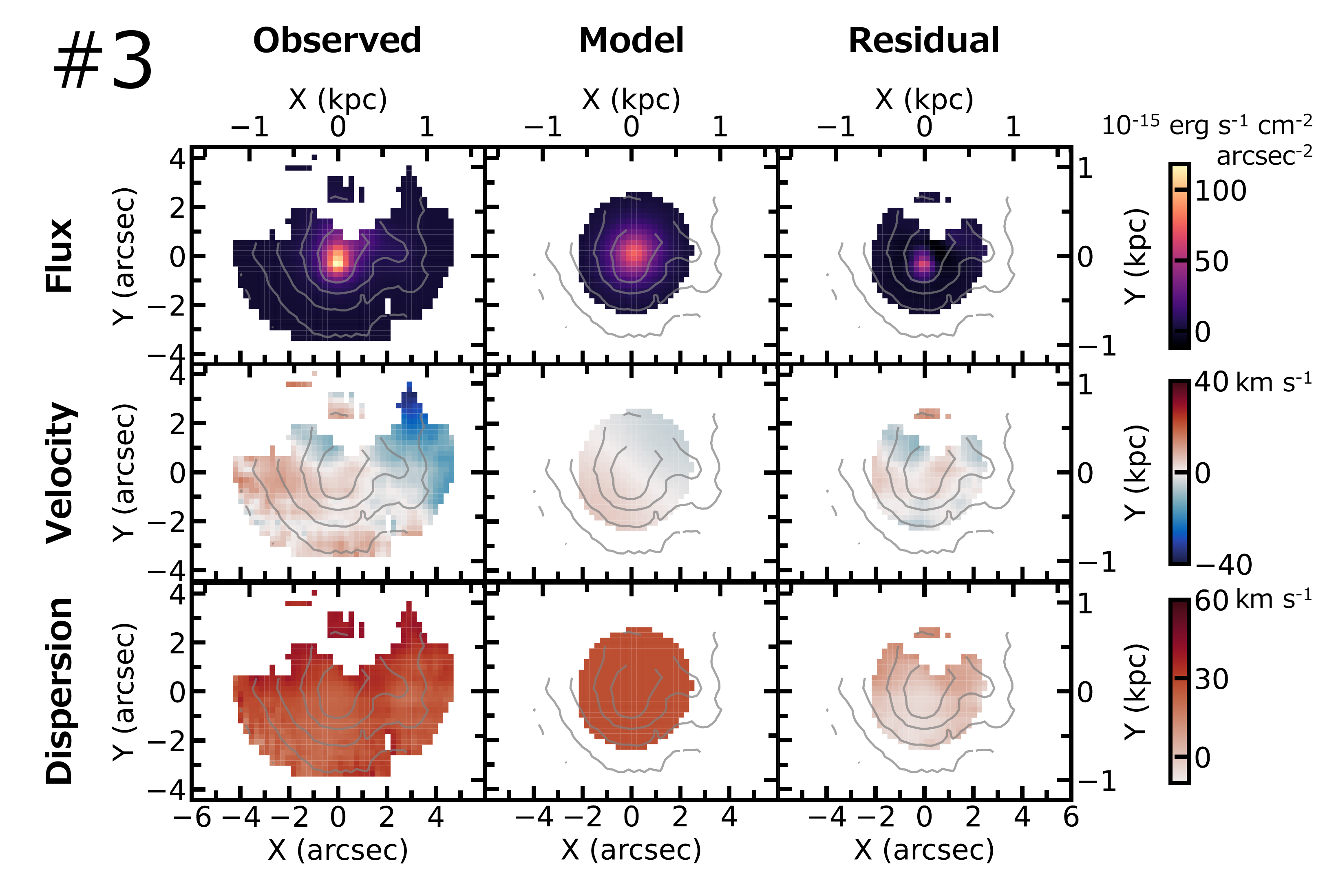}
    \caption{GalPaK$^{\rm 3D}$ result of SBS0335$-$052E.}
    \label{fig:galpak_3}
\end{figure*}

\begin{figure*}[t]
    \centering
    \includegraphics[width=18.0cm]{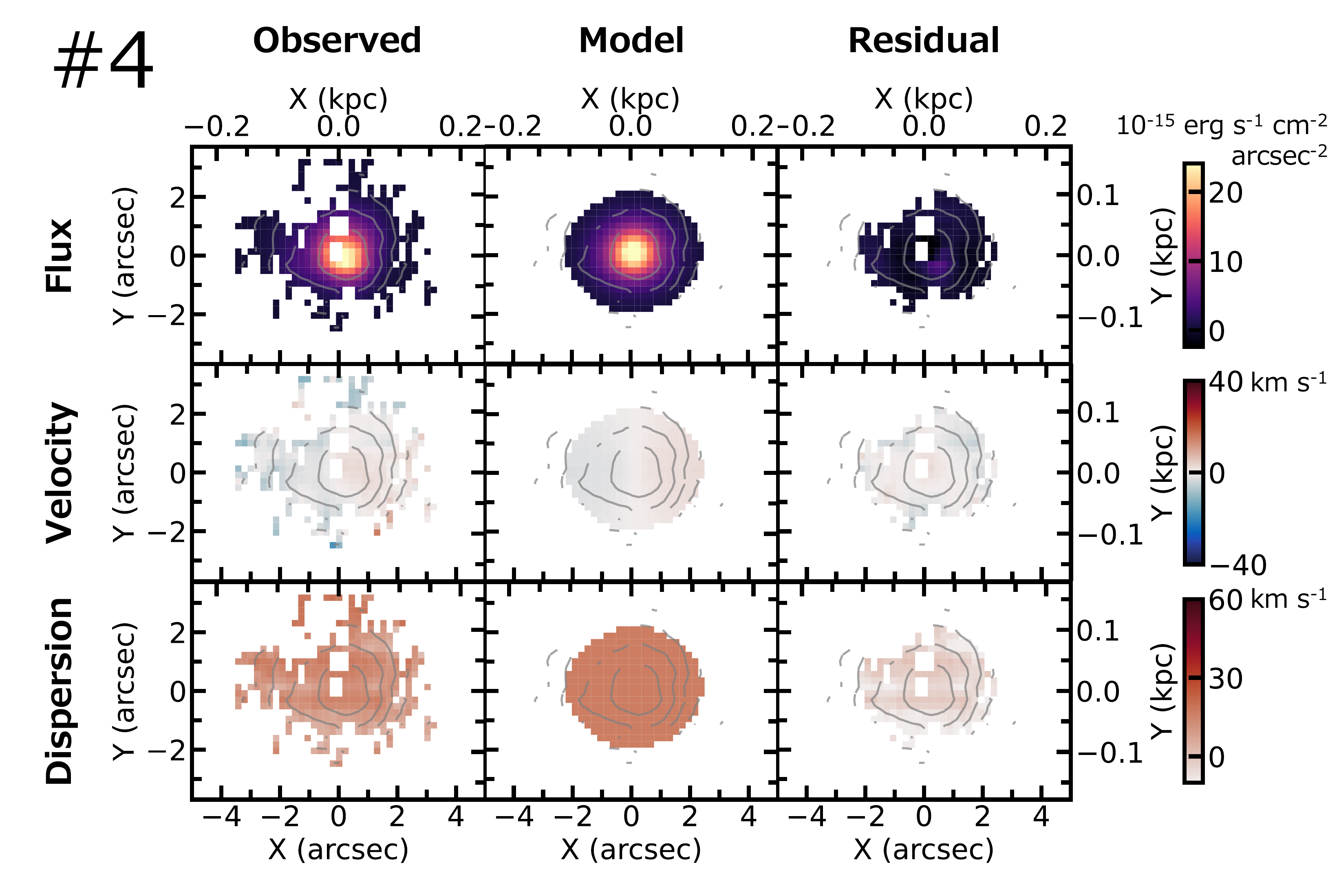}
    \caption{GalPaK$^{\rm 3D}$ result of HS0822+3542.}
    \label{fig:galpak_4}
\end{figure*}

\begin{figure*}[t]
    \centering
    \includegraphics[width=18.0cm]{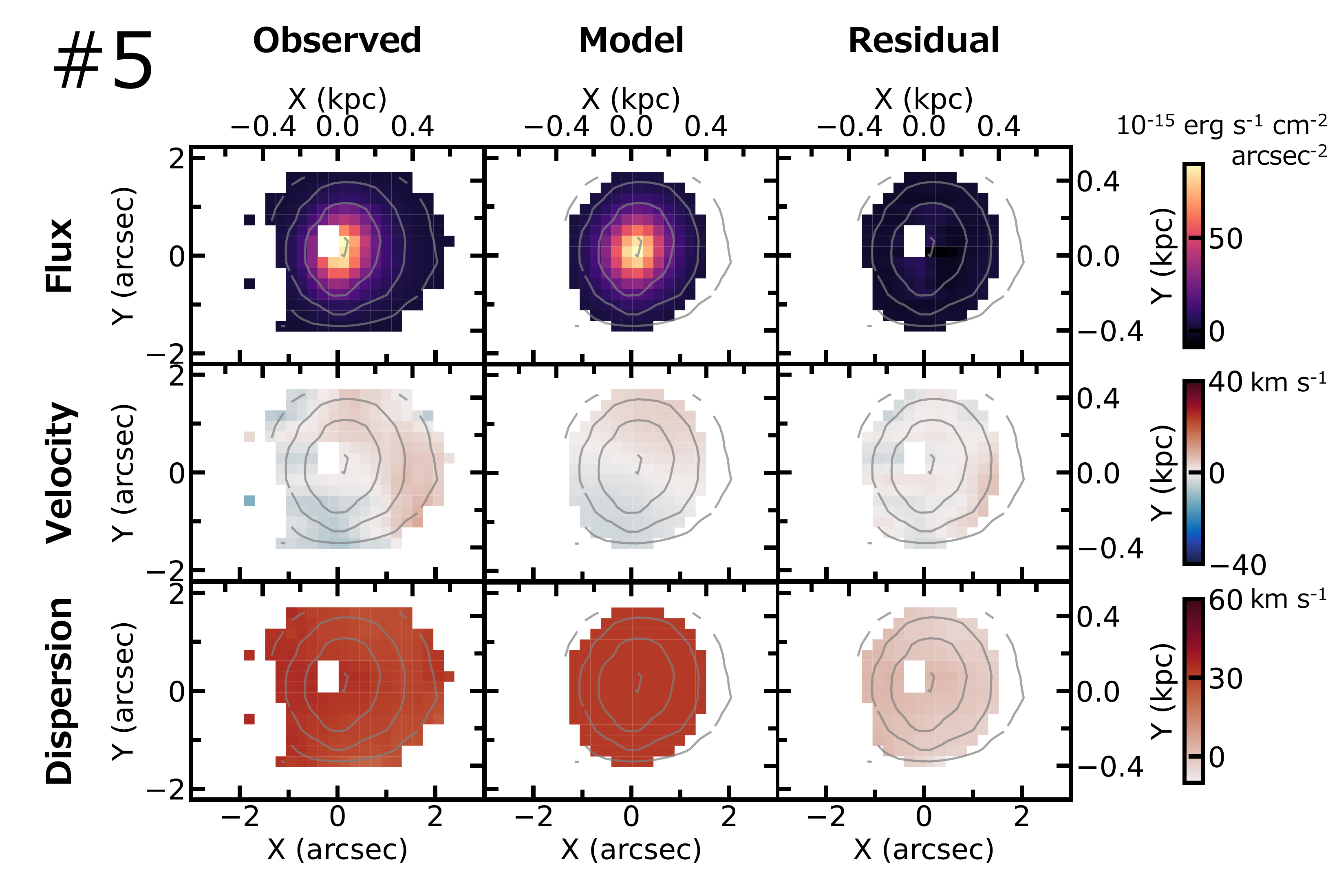}
    \caption{GalPaK$^{\rm 3D}$ result of J1044+0353.}
    \label{fig:galpak_5}
\end{figure*}

\begin{figure*}[t]
    \centering
    \includegraphics[width=18.0cm]{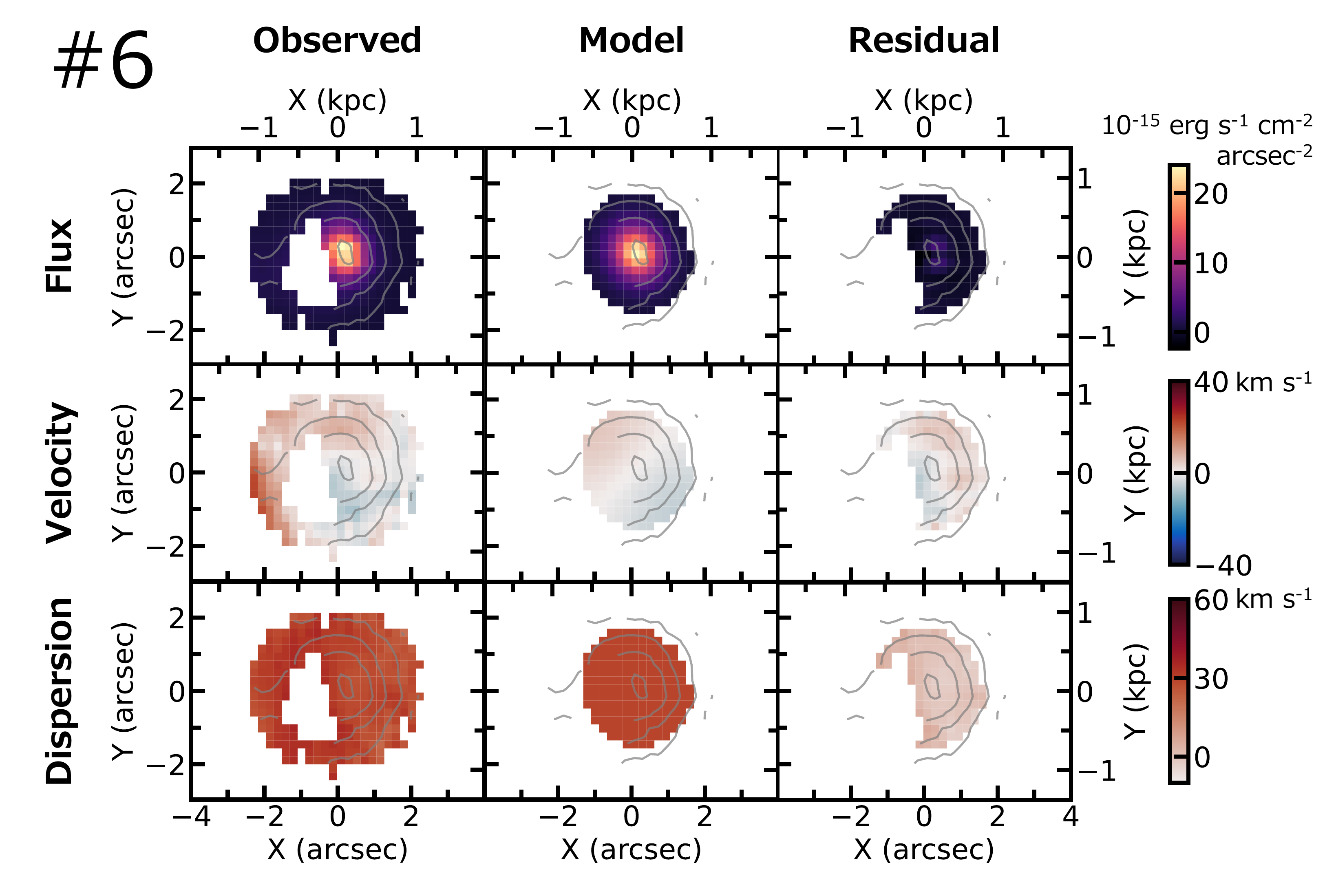}
    \caption{GalPaK$^{\rm 3D}$ result of J2115$-$1734.}
    \label{fig:galpak_6}
\end{figure*}

\tcrb{Figures \ref{fig:galpak_2}--\ref{fig:galpak_6} show GalPaK$^{\rm 3D}$ fitting results of the 5 EMPGs other than J1631+4426.}

%

\bibliography{library}


\end{document}